\documentclass[a4paper,11pt]{article}

\usepackage{jheppub} 

\usepackage[T1]{fontenc} 
\usepackage{graphicx}
\usepackage{bm}
\usepackage{amsmath}
\usepackage{amssymb}
\usepackage{subfigure}
\usepackage[mathlines]{lineno}

\title{\boldmath Improved measurements of the neutrino mixing angle $\theta_{13}$ with the Double Chooz detector}

\newcommand{\Aachen}{III. Physikalisches Institut, RWTH Aachen University, 52056 Aachen, Germany}
\newcommand{\Alabama}{Department of Physics and Astronomy, University of Alabama, Tuscaloosa, Alabama 35487, USA}
\newcommand{\Argonne}{Argonne National Laboratory, Argonne, Illinois 60439, USA}
\newcommand{\APC}{AstroParticule et Cosmologie, Universit\'{e} Paris Diderot, CNRS/IN2P3, CEA/IRFU, Observatoire de Paris, Sorbonne Paris Cit\'{e}, 75205 Paris Cedex 13, France}
\newcommand{\CBPF}{Centro Brasileiro de Pesquisas F\'{i}sicas, Rio de Janeiro, RJ, 22290-180, Brazil}
\newcommand{\Chicago}{The Enrico Fermi Institute, The University of Chicago, Chicago, Illinois 60637, USA}
\newcommand{\CIEMAT}{Centro de Investigaciones Energ\'{e}ticas, Medioambientales y Tecnol\'{o}gicas, CIEMAT, 28040, Madrid, Spain}
\newcommand{\Columbia}{Columbia University; New York, New York 10027, USA}
\newcommand{\Davis}{University of California, Davis, California 95616, USA}
\newcommand{\Drexel}{Department of Physics, Drexel University, Philadelphia, Pennsylvania 19104, USA}
\newcommand{\Hiroshima}{Hiroshima Institute of Technology, Hiroshima, 731-5193, Japan}
\newcommand{\IIT}{Department of Physics, Illinois Institute of Technology, Chicago, Illinois 60616, USA}
\newcommand{\INR}{Institute of Nuclear Research of the Russian Academy of Sciences, Moscow 117312, Russia}
\newcommand{\CEA}{Commissariat \`{a} l'Energie Atomique et aux Energies Alternatives, Centre de Saclay, IRFU, 91191 Gif-sur-Yvette, France}
\newcommand{\Kansas}{Department of Physics, Kansas State University, Manhattan, Kansas 66506, USA}
\newcommand{\Kobe}{Department of Physics, Kobe University, Kobe, 657-8501, Japan}
\newcommand{\Kurchatov}{NRC Kurchatov Institute, 123182 Moscow, Russia}
\newcommand{\MIT}{Massachusetts Institute of Technology, Cambridge, Massachusetts 02139, USA}
\newcommand{\MaxPlanck}{Max-Planck-Institut f\"{u}r Kernphysik, 69117 Heidelberg, Germany}
\newcommand{\Niigata}{Department of Physics, Niigata University, Niigata, 950-2181, Japan}
\newcommand{\NotreDame}{University of Notre Dame, Notre Dame, Indiana 46556, USA}
\newcommand{\IPHC}{IPHC, Universit\'{e} de Strasbourg, CNRS/IN2P3, 67037 Strasbourg, France}
\newcommand{\SUBATECH}{SUBATECH, CNRS/IN2P3, Universit\'{e} de Nantes, Ecole des Mines de Nantes, 44307 Nantes, France}
\newcommand{\Tennessee}{Department of Physics and Astronomy, University of Tennessee, Knoxville, Tennessee 37996, USA}
\newcommand{\TohokuUni}{Research Center for Neutrino Science, Tohoku University, Sendai 980-8578, Japan}
\newcommand{\TohokuGakuin}{Tohoku Gakuin University, Sendai, 981-3193, Japan}
\newcommand{\TokyoInst}{Department of Physics, Tokyo Institute of Technology, Tokyo, 152-8551, Japan }
\newcommand{\TokyoMet}{Department of Physics, Tokyo Metropolitan University, Tokyo, 192-0397, Japan}
\newcommand{\Muenchen}{Physik Department, Technische Universit\"{a}t M\"{u}nchen, 85748 Garching, Germany}
\newcommand{\Tubingen}{Kepler Center for Astro and Particle Physics, Universit\"{a}t T\"{u}bingen, 72076 T\"{u}bingen, Germany}
\newcommand{\UFABC}{Universidade Federal do ABC, UFABC, Santo Andr\'{e}, SP, 09210-580, Brazil}
\newcommand{\UNICAMP}{Universidade Estadual de Campinas-UNICAMP, Campinas, SP, 13083-970, Brazil}
\newcommand{\vtech}{Center for Neutrino Physics, Virginia Tech, Blacksburg, Virginia 24061, USA}

\author{Double Chooz Collaboration}
\author[aa]{\\Y.~Abe,}
\author[e]{J.C.~dos Anjos,}
\author[n]{J.C.~Barriere,}
\author[v]{E.~Baussan,}
\author[a]{I.~Bekman,}
\author[i]{M.~Bergevin,}
\author[y]{T.J.C.~Bezerra,}
\author[m]{L.~Bezrukov,}
\author[f]{E.~Blucher,}
% ...

\affiliation[a]{\Aachen}
% ...
\affiliation[e]{\CBPF}
\affiliation[f]{\Chicago}
% ...
\affiliation[i]{\Davis}
% ...
\affiliation[m]{\INR}
\affiliation[n]{\CEA}
% ...
\affiliation[v]{\IPHC}
% ...
\affiliation[y]{\TohokuUni}
% ...
\affiliation[aa]{\TokyoInst}
% ...

\emailAdd{kuze@phys.titech.ac.jp}

\abstract
{The Double Chooz experiment presents improved measurements of the neutrino mixing angle $\theta_{13}$ using the data collected in 467.90 live days from a detector positioned at an average distance of 1050\,m from two reactor cores at the Chooz nuclear power plant.
Several novel techniques have been developed to achieve significant reductions of the backgrounds and systematic uncertainties with respect to previous publications, whereas the efficiency of the $\bar\nu_{e}$ signal has increased.
The value of $\theta_{13}$ is measured to be $\sin^{2}2\theta_{13} = 0.090 ^{+0.032}_{-0.029}$ from a fit to the observed energy spectrum.
Deviations from the reactor $\bar\nu_{e}$ prediction observed above a prompt signal energy of 4\,MeV and possible explanations are also reported.
A consistent value of $\theta_{13}$ is obtained from a fit to the observed rate as a function of the reactor power independently of the spectrum shape and background estimation, demonstrating the robustness of the $\theta_{13}$ measurement despite the observed distortion.}

\begin{document}
\maketitle
\flushbottom

\section{Introduction}

In the standard three-flavor framework, the neutrino oscillation probability is determined by three mixing angles, three mass-squared differences (of which two are independent) and one CP-violation phase.
Among the three mixing angles, $\theta_{13}$ has been measured recently by $\bar\nu_e$ disappearance in short-baseline experiments~\cite{ref:DCII_nGd, ref:DCII_nH, ref:DCII_RRM, ref:DayaBay, ref:RENO} and $\nu_{\mu} \rightarrow \nu_{e}$ appearance in long-baseline experiments~\cite{ref:MINOS_t13, ref:T2K_t13}.
The other two angles had been measured before~\cite{ref:PDG}, while the mass hierarchy of neutrinos and CP-violation phase are still unknown.
The discovery potential of future projects critically depends on the values of the mixing angles and, therefore, a precise measurement of $\theta_{13}$ is essential for deep understandings of neutrino physics expected in the following decades.

According to the current knowledge, one mass-squared difference is much smaller than the other.
This allows us to interpret the experimental data by a simple two-flavor oscillation scheme in many cases.
In the two-flavor scheme, survival probability of $\bar\nu_e$ with energy $E_{\nu}$\,(MeV)  after traveling a distance of $L$\,(m) is expressed as:
\begin{equation}
P = 1 - \sin^{2}2\theta_{13}\sin^2\left(1.27\,\Delta m_{31}^{2} {\rm (eV^{2})}\,L/E_{\nu}\right).
\label{eq:oscillation}
\end{equation}
This equation is a good approximation to reactor neutrino oscillation for $L$ less than a few km, and the matter effect is negligible as well.
Therefore, the value of $\theta_{13}$ can be directly measured from the oscillation amplitude in reactor neutrino oscillation.

Reactor neutrinos are detected by a delayed coincidence technique through the inverse $\beta$-decay (IBD) reaction on protons: $\bar\nu_e + p \rightarrow e^{+} + n$.
The positron is observed as the prompt signal with the energy related to the neutrino energy as: $E_{\rm signal} \simeq E_{\nu} - 0.8\,{\rm MeV}$.
The neutron is captured either on Gd or H in liquid scintillator with high efficiency.
Gd captures occur after a mean time of 31.1\,$\mu$s and emit a few $\gamma$-rays with a total energy of 8\,MeV, which is well above the energy of natural radioactivity and easily distinguishable from the random coincidence of such background.
Double Chooz has developed a new analysis of $\bar\nu_e$ disappearance measurement using a coincidence with H captures~\cite{ref:DCII_nH}, but these additional captures are not used in the analysis presented in this paper.

Here we report on improved measurements of $\theta_{13}$ using the data collected by the Double Chooz far detector (FD) in 467.90 live days with 66.5\,GW-ton-years of exposure (reactor power $\times$ detector mass $\times$ live time), corresponding to a factor of two more statistics compared to the previous publication~\cite{ref:DCII_nGd}.
The analysis is based on a new method of energy reconstruction described in Section~\ref{section:Reconstruction}.
After the delayed coincidence is required, remaining backgrounds are mostly induced by cosmic muons, including long-lived cosmogenic isotopes, proton recoils by muon-induced spallation neutrons and stopping muons.
Several novel techniques have been developed in the new analysis to suppress such backgrounds (Section~\ref{section:Selection}).
In contrast, IBD signal efficiency has increased with the extended signal window and, together with the newly developed analysis methods, detection systematic uncertainties have been reduced by almost a factor of two with respect to the previous analysis (Section~\ref{section:DetectionEfficiency}).
Remaining backgrounds are estimated by dedicated studies described in Section~\ref{section:Background} and also directly measured in reactor-off running as shown in Section~\ref{section:ReactorOff}.
The value of $\theta_{13}$ is extracted from a fit to the prompt energy spectrum.
Additional deviations from the reactor $\bar\nu_{e}$ prediction are observed above 4\,MeV although the impact on the $\theta_{13}$ measurement is not significant.
A consistent value of $\theta_{13}$ is also obtained by a fit to the observed rates as a function of reactor power, which provides a complementary measurement independent of the energy spectrum shape and background estimation.
Results of the neutrino oscillation analyses and investigation of observed spectral distortion are discussed in Section~\ref{section:Oscillation} and \ref{section:Distortion}.

In the current analysis, with only the far detector, the precision of $\theta_{13}$ measurement is limited by the systematic uncertainty on the flux prediction.
After the cancellation of the flux and other systematic uncertainties using the near detector (ND), which is currently under construction, uncertainties on the background should be dominant.
Improvements of the analysis described in this paper are therefore critical to enhance the sensitivity of the future Double Chooz data taken with the ND.
The projected sensitivity is studied based on the improved analysis in Section~\ref{section:ProjectedSensitivity}.

\section{Experimental setup}
\label{section:ExperimentalSetup}
The Double Chooz far detector is located at an average distance of 1,050\,m from the two reactor cores, in a hill topology with 300 meters water equivalent (m.w.e.) rock overburden to shield cosmic muons.
In this section, we briefly review the detector and the Monte Carlo simulation.
More details are given elsewhere~\cite{ref:DCII_nGd}.

\subsection{Double Chooz Detector}
\label{section:Detector}
\begin{figure}
\begin{center}
	\includegraphics[width=12cm]{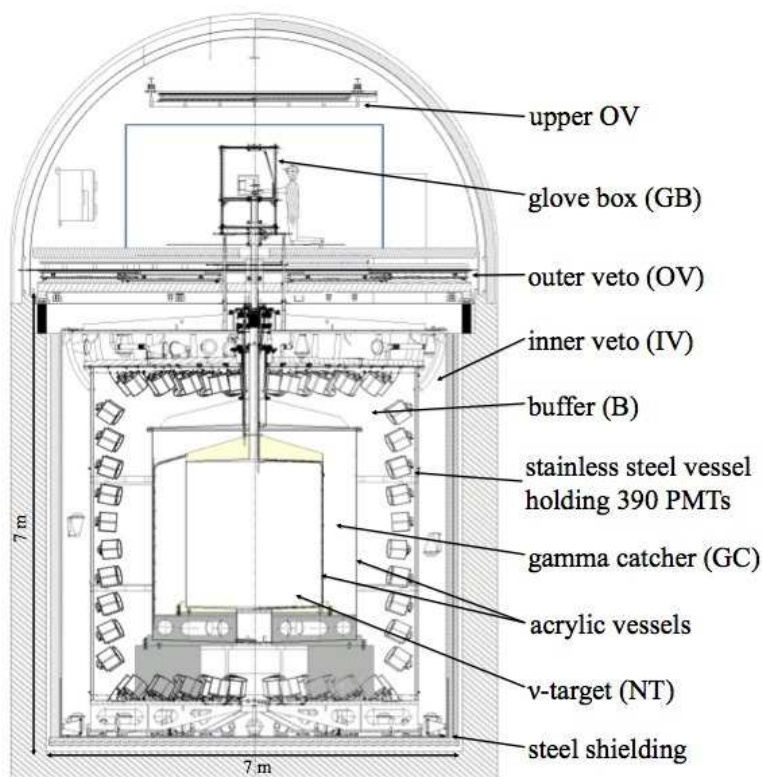}
	\caption{Schematic view of the Double Chooz detector.}
	\label{fig:Detector}
\end{center}
\end{figure}

Double Chooz has developed a calorimetric liquid scintillator detector made of four concentric cylindrical vessels optimized for detection of reactor neutrinos.
Figure~\ref{fig:Detector} shows a schematic view of the Double Chooz detector.
The innermost volume, named $\nu$-target (NT), is filled with 10\,m$^3$ Gd-loaded liquid scintillator~\cite{ref:GdScintillator}.
NT is surrounded by a 55\,cm thick Gd-free liquid scintillator layer, called the $\gamma$-catcher (GC).
When neutrons from IBD interactions are captured on Gd in the NT, $\gamma$-rays with a total energy of 8\,MeV are emitted.
These $\gamma$-rays are detected either by the NT and/or the GC.
The GC is further surrounded by a 105\,cm thick non-scintillating mineral oil layer, called the Buffer.
The boundaries of the NT, GC and Buffer are made of transparent acrylic vessels, while the Buffer volume is surrounded by a steel tank and optically separated from an outer layer described below.
There are 390 low background 10-inch photomultiplier tubes (PMTs)~\cite{ref:pmt,ref:pmt2} positioned on the inner surface of the buffer tank.
Orientation and positions of the PMT assemblies and dimensions of the tank walls were verified by photographic survey.
The Buffer layer works as a shield to $\gamma$-rays from radioactivity of  PMTs and the surrounding rock.
The inner three regions and PMTs are collectively referred to as the inner detector (ID).
Outside of the ID is a 50 cm thick liquid scintillator layer called the inner veto (IV).
The IV is equipped with 78 8-inch PMTs, among which 24 PMTs are arranged on the top, 12 PMTs at mid-height on the side walls and 42 PMTs on the bottom.
The IV works not only as an active veto to cosmic ray muons but as a shield to fast neutrons from outside of the detector.
Fast neutrons are often tagged by the IV as well.
The whole detector is further surrounded by a 15 cm thick steel shield to protect it against external $\gamma$-rays.
Each ID and IV PMT is surrounded by mu-metal to suppress magnetic field from the Earth and the steel shield~\cite{ref:pmt3}. 
A central chimney, connected to all layers, allows the introduction of the liquids and calibration sources. 

Digitized signal waveforms from all ID and IV PMTs are recorded by 8-bit flash-ADC electronics with 500\,MHz sampling~\cite{ref:fadc}.
The trigger threshold is set at 350\,keV, well below the minimum energy of $\bar\nu_{e}$ signal at 1.02\,MeV.
 The energy threshold is lowered from 500\,keV to 400\,keV in the new analysis  but the trigger efficiency still reaches 100\,\% with negligible uncertainty.

An outer veto (OV) covers the top of the detector tank.
The OV consists of plastic scintillator strips with cross-section of 5\,cm $\times$ 1\,cm.
Two layers with orthogonally oriented strip directions cover a 13\,m $\times$ 7\,m area except for around the chimney, and another two layers are mounted above the chimney region.
The full OV has been installed for 27.6\,\% of the data presented in this paper, while only the lower two layers have been available for 56.7\,\% including the data taken with reactor off (see Section~\ref{section:ReactorOff}).
The remaining 15.7\,\% of data have been recorded without OV.

Double Chooz developed several calibration systems to suppress systematic uncertainties associated with the detector response.
A laser system is used to illuminate ID PMTs through an optical fiber and diffuser ball deployed inside the NT.
The time offset for each readout channel is measured by the laser system with an uncertainty of 0.15\,ns.
In addition, multi-wavelength LED-fiber systems are used to inject light into the ID and IV from a set of fixed points mounted on the PMT covers.
The data with the LED-fiber systems are taken regularly to measure the characteristics of the readout electronics, i.e. time offset and gain, and their stability during the operation.
Radioisotopes, $^{68}$Ge, $^{137}$Cs, $^{60}$Co and $^{252}$Cf have been deployed in the NT along the vertical symmetry axis and in the GC through a tube along the boundaries to the NT and Buffer.
Among them, $^{252}$Cf  neutron source is used to evaluate systematic uncertainties on the detection efficiency of neutron captures on Gd.
In addition to these calibration devices, abundant spallation neutrons captured on H, Gd and C and Bi-Po decays from radio-impurity in the liquid scintillator are used for various calibration purposes.
These events are distributed over the detector volume and constantly observed during data taking, and therefore suitable for extracting corrections for the time stability and position dependence of the energy scale.

\subsection{Reactor and Detector Models}
\label{section:Simulation}
Expectation of reactor $\bar\nu_{e}$ events in the Double Chooz detector is calculated by a Monte Carlo (MC) simulation.
The MC has two constituents: IBD interaction of reactor $\bar\nu_{e}$ in the detector and simulation of the detector response.
Neutrino oscillation is studied by comparing the observed IBD candidates with the prediction from the MC.
In addition, MC data with radioactive sources are generated and used to evaluate the systematic uncertainties on the energy scale and detection efficiency.
However, the background rate and energy spectrum are estimated with the data and the MC is used only for validation.

\subsubsection{Reactor $\bar\nu_{e}$ prediction}
\label{section:FluxPrediction}
Double Chooz observes $\bar\nu_e$ from the two reactor cores at Chooz nuclear power plant operated by \'{E}lectricit\'{e} de France (EDF), both with 4.25\,GW$_{\rm th}$ thermal power.
The instantaneous thermal power of each reactor core is provided by EDF with time steps of $<$ 1 minute with an uncertainty of 0.5\,\% at the full reactor power.
$\bar\nu_e$'s are produced in nuclear reactors by the $\beta$-decay of fission products, in which more than 99.7\,\% of fissions originate from four isotopes, $^{235}$U,  $^{239}$Pu,  $^{238}$U, and $^{241}$Pu.
The reference $\bar\nu_{e}$ spectra are derived for $^{235}$U,  $^{239}$Pu and $^{241}$Pu from fits to their $\beta$ spectrum measured at the ILL research reactor~\cite{ref:SchreckU5,ref:SchreckU5Pu9,ref:SchreckPu9Pu1} considering the allowed transitions.
It was highlighted that contribution of the forbidden transitions have some impact on the neutrino spectrum~\cite{ref:Hayes}, but this is under investigation.
In addition, a measurement of the  spectrum from $^{238}$U~\cite{ref:U238} is newly used in this analysis with an extrapolation below 3\,MeV and above 7.5\,MeV, using a combination of the summation method~\cite{ref:Summation} and an exponential-polynomial fit on the data. 
In the previous analysis, the $^{238}$U contribution was derived from a calculation~\cite{ref:238Calc}.
Estimation of $\bar\nu_{e}$ flux based on the new measurement is larger by 10\,\% at 3\,MeV with respect to the previous calculation for the $^{238}$U contribution, which is itself roughly 10\,\% of the total $\bar\nu_{e}$ flux. 
Evolution of each fractional fission rate and associated errors are evaluated using a full core model and assembly simulations developed with the MURE code~\cite{ref:MURE, ref:MURE-NEA}. 
Benchmarks with other codes have been performed~\cite{ref:Takahama} in order to validate the simulations.
Locations and initial burn-up of each assembly are provided by EDF  for every reactor fuel cycle with approximately one year in duration and used as input to the core simulation.
In order to suppress the normalization uncertainty in the $\bar\nu_e$ prediction, Double Chooz used the $\bar\nu_e$ rate measurement by Bugey4~\cite{ref:Bugey4} at a distance of 15\,m with corrections for the different fuel composition between Bugey4 and each of the Chooz cores.
Systematic uncertainty on the IBD signal rate associated with the flux prediction is evaluated to be 1.7\,\% of which the dominant component is an uncertainty of 1.4\,\% in the Bugey4 measurement as shown in Table~\ref{table:FluxError}, while the uncertainty would have been 2.8\,\% without use of the Bugey4 rate measurement.
Contributions from forbidden $\beta$ decays are effectively integrated in the Bugey4 rate measurement, whereas it neglects possible influence to the spectrum shape.

\begin{table}[ht]
  \footnotesize
  \begin{center}
    \begin{tabular}{| l | c | }
      \hline
      Source & Uncertainty (\%)\\
      \hline
      Bugey4 measurement & $1.4$ \\
      Fractional fission rate of each isotope & $0.8$ \\
      Thermal power & $0.5$\\
      IBD cross-section & $0.2$ \\
      Mean energy released per fission & $0.2$ \\
      Distance to reactor core & $< 0.1$ \\
      \hline
      Total & 1.7 \\
      \hline
    \end{tabular}
  \end{center}
  \caption{Uncertainties on the reactor $\bar\nu_{e}$ rate prediction. Uncertainties on the energy spectrum shape are also accounted for in the neutrino oscillation analysis.}
  \label{table:FluxError}
\end{table}

\subsubsection{Detector simulation}
\label{section:DetectorSimulation}
Double Chooz developed a detector simulation based on Geant4.9.2.p02~\cite{ref:Geant4_1, ref:Geant4_2} with custom models of the neutron scattering, Gd $\gamma$ cascade, scintillation processes and photocathode optical surface.
A custom neutron scattering model implements hydrogen molecular bonds in elastic scattering below 4\,eV based on Ref.~\cite{ref:NeutronTH} and an improved radiative capture model below 300\,eV.
It was confirmed that the MC simulation with the custom simulation code reproduces the observed neutron capture time better than the default Geant4, especially for short capture time which is sensitive to the thermalization process.
The detector geometry is implemented in the simulation including the acrylic and steel vessels, support structure, PMTs and mu-metal shields.
Optical parameters of liquids including the light yield of NT and GC liquid scintillators, photoemission time probabilities, light attenuation and ionization quenching treatments are based on lab measurements.

Double Chooz also developed a custom readout simulation, which accounts for the response of the full readout system including the PMT, front-end electronics, flash-ADC, trigger system and data acquisition.
The simulation implements a probability distribution function to empirically characterize the response to each single photoelectron (p.e.) based on measurements.
Single p.e.'s are accumulated to produce the waveform signal for each PMT, and the waveform is digitized by the flash-ADC conversion with a 2\,ns time bin.
Channel-to-channel variations of the readout response such as gains, baselines and noise are taken into account to accurately predict resolution effects.
Reactor $\bar\nu_e$ events are generated over the detector volume by the full MC simulation and compared with the data.

\section{Event Reconstruction}
\label{section:Reconstruction}

\subsection{FADC Pulse Reconstruction}
\label{section:PulseReconstruction}
Event reconstruction starts from pulse reconstruction, which extracts the signal charge and time for each PMT from the digitized waveform recorded by the flash-ADC.
Periodic triggers are taken with a rate of 1\,Hz for the full readout time window (256\,ns) in order to compute the mean ADC counts of the baseline, $B_{\rm mean}$, and its fluctuation as RMS, $B_{\rm rms}$, for each readout channel.
The integrated charge is defined as the sum of ADC counts during the integration time window after $B_{\rm mean}$ is subtracted.
The length of the integration time window (112\,ns) is chosen to optimize the charge resolution of single p.e. signal, energy resolution and charge integration efficiency. 
The start time of the integration time window is determined to maximize the integrated charge for each channel for each event.
For events depositing up to a few MeV in the NT, most PMTs detect only one p.e., which typically has an amplitude of about 6 ADC counts.
In order to discriminate against noise fluctuations in the absence of an actual p.e. signal, the following conditions are required: $\ge 2$ ADC counts in the maximum bin and $q > B_{\rm rms} \times \sqrt{N_{\rm s}}$ where $q$ is the integrated charge and $N_{\rm s}$ is the number of samples in the integration window (56 for a 112\,ns window).
Charge and time in the MC simulation are extracted from digitized waveforms given by the readout simulation following the same procedure as that for data.

\subsection{Event Vertex Reconstruction}
\label{section:EventVertexReconstruction}
The vertex position of each event is reconstructed based on a maximum likelihood algorithm using charge and time, assuming the event to be a point-like light source.
The event likelihood is defined as:
\begin{equation}
\mathcal{L}(\mathbf{X}) = \prod_{q_{i}=0}f_{q}(0; q_{i}')\prod_{q_{i}>0}f_{q}(q_{i}; q_{i}')f_{t}(t_{i};t_{i}',q_{i}'),
\label{eq:vertex}
\end{equation}
where $q_{i}$ and $t_{i}$ are the observed charge and time for the $i$-th readout channel, respectively.
$q_{i}'$ and $t_{i}'$ are the expected charge and time for each channel from a point-like light source with the position, time and light intensity per unit solid angle ($\Phi$) given by $\mathbf{X} = (x, y, z, t, \Phi)$.
$f_{q}$ and $f_{t}$ are the probability to measure the charge and time given the predictions.
The best possible set of $\mathbf{X}$ is found by maximizing $\mathcal{L}(\mathbf{X})$, which is equivalent to minimizing the negative log-likelihood function:
\begin{equation}
F_{\rm V} = -\ln{\mathcal{L}(\mathbf{X})}.
\label{eq:FuncV}
\end{equation}
Effective light attenuation and PMT angular response used in the event vertex reconstruction are tuned using source calibration data, and the charge and time likelihoods are extracted from laser calibration data.
Both the performance of the event vertex reconstruction and agreement between the data and MC are improved with the tuning.

\subsection{Energy Reconstruction}
\label{section:EnergyReconstruction}
Visible energy, $E_{\rm vis}$, is reconstructed from the total number of photoelectrons, $N_{\rm pe}$, as:
\begin{eqnarray}
E_{\rm vis} &=& N_{\rm pe}
\times f^{\rm data}_{\rm u}(\rho, z) \times f^{\rm data}_{\rm MeV}  \times f_{\rm s}(E_{\rm vis}^{0}, t) \quad \mbox{for the data}
\label{eq:EnergyData}
\end{eqnarray}
and
\begin{eqnarray}
E_{\rm vis} &=& N_{\rm pe}
\times f^{\rm MC}_{\rm u}(\rho, z) \times f^{\rm MC}_{\rm MeV} \times f_{\rm nl}(E_{\rm vis}^{0}) \quad \mbox{for the MC.}
\label{eq:EnergyMC}
\end{eqnarray}
The parameters $\rho$ and $z$ represent the vertex position in the detector coordinate with $\rho$ the radial distance from the central vertical axis and $z$ the vertical coordinate and $t$ is the event time (elapsed days).
Corrections for the uniformity ($f_{\rm u}$), absolute energy scale ($f_{\rm MeV}$), time stability ($f_{\rm s}$) and non-linearity ($f_{\rm nl}$) are applied to get the final visible energy.
$E_{\rm vis}^{0}$ represents the energy after applying the uniformity correction, which is subsequently subject to the energy-dependent corrections for the stability and non-linearity.
Visible energy from the MC simulation is obtained following the same procedure as that for the data, although the stability correction is applied only to the data and the non-linearity correction is applied only to the MC.
Each correction is explained in the following subsections.

\subsubsection{Linearized PE calibration}
\label{section:LinearizedPE}
\begin{figure}
\begin{center}
	\includegraphics[width=8cm]{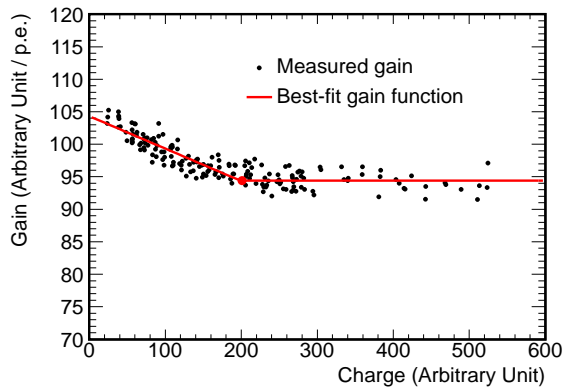}
	\caption{Gain as a function of integrated charge for a typical readout channel. Points show the measurements and the line shows the gain function obtained from a fit with three parameters explained in the text.}
	\label{fig:LinearPE}
\end{center}
\end{figure}

The total number of photoelectrons is given as $N_{\rm pe} 
= \sum_{i}q_{i}/g^{m}_{i}(q_{i},t)$ where $i$ refers to each readout channel and $m$ refers to either $\rm data$ or $\rm MC$.
$q_{i}$ is the integrated charge by the pulse reconstruction and $g^{m}_{i}$ is a charge-to-p.e. conversion factor (referred to as gain) extracted by calibration taking into account the variation in the course of the data taking (elapsed days, $t$) and charge dependence, i.e. gain non-linearity.
Due to limited sampling of the waveform digitizer, the baseline estimation can be biased within $\pm$1 ADC count, which results in a gain non-linearity especially below a few photoelectrons~\cite{ref:FADC_NL}.
Gain is measured using the data taken with a constant light yield provided by the LED-fiber calibration systems, as $g_{i} = \alpha \times \sigma^{2}_{i}/\mu_{i}$, where $\mu_{i}$ and $\sigma_{i}$ are the mean and standard deviation (RMS) of the observed charge distribution.
The parameter $\alpha$ is used to correct for the intrinsic spread in $\sigma_{i}$ due to single p.e. width and electronic noise.
It is considered to be constant for all readout channels and is chosen by making the number of photoelectrons in the H capture of spallation neutrons equal to the hit PMT multiplicity ($n$). 
Non-single p.e. contributions are taken into account using Poisson statistics as: $n = -N_{\rm PMT} \ln{(1 - N_{\rm hits}/N_{\rm PMT})}$, where $N_{\rm PMT}$ and $N_{\rm hits}$ are the number of all PMTs and hit PMTs, respectively.
Calibration data are taken with different light intensities and light injection positions to measure the gain non-linearity of all channels.
Figure~\ref{fig:LinearPE} shows the measured gain as a function of integrated charge for a typical readout channel, overlaid with the gain correction function characterized by three parameters: constant gain at high charge, non-linearity slope at low charge and the transition point.
Since the gain and its non-linearity change after power cycles of the readout electronics, the gain is measured upon each power-cycle period.
Time dependence during each power-cycle period is further corrected using natural calibration sources as described in a later section.

\subsubsection{Uniformity calibration}
\label{section:Uniformity}
\begin{figure}
\begin{center}
	\includegraphics[width=8cm]{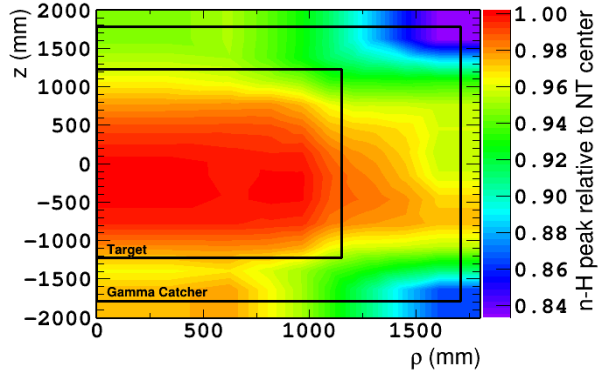}
	\caption{Uniformity correction map for the data obtained by fitting the neutron capture peak on hydrogen.}
	\label{fig:CorrectionMap}
\end{center}
\end{figure}

Uniformity calibration is introduced to correct for the position dependence of $N_{\rm pe}$.
A correction is applied as a function of $\rho$ and $z$ to convert $N_{\rm pe}$ into that at the center of the detector.
The function $f_{\rm u}(\rho, z)$ for the data is obtained as shown in Fig.~\ref{fig:CorrectionMap} using $\gamma$'s from neutron captures on H, which peak at 2.2\,MeV.
The correction factor ranges up to around 5\,\%  inside the NT.
A similar pattern is seen in the correction map for the MC.
The systematic uncertainty due to the non-uniformity of the energy scale is evaluated to be 0.36\,\% from the residual position-dependent differences between data and MC measured with $\gamma$'s from neutron captures on Gd.

\subsubsection{Energy scale calibration}
\begin{figure}
\begin{center}
	\includegraphics[width=8cm]{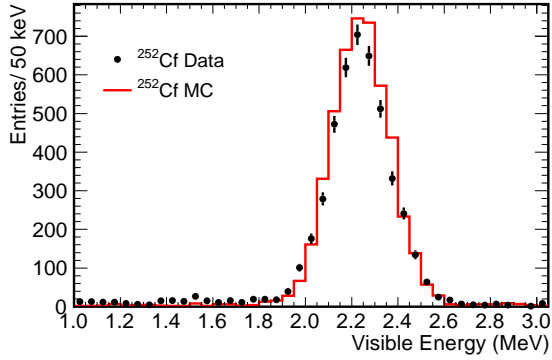}
	\caption{Neutron capture peak on hydrogen: points show the data taken with $^{252}$Cf neutron source deployed at the center of the detector and the histogram shows the corresponding MC simulation.}
	\label{fig:AbsoluteEnergy}
\end{center}
\end{figure}

The absolute energy scale is determined by the position of the 2.223\,MeV peak of neutrons captured on H using the data taken with a $^{252}$Cf neutron source deployed at the center of the detector (Fig.~\ref{fig:AbsoluteEnergy}).
The absolute energy scale ($1/f_{\rm MeV}$) is found to be 186.2\,p.e./MeV for the data and 186.6\,p.e./MeV for the MC.

\subsubsection{Stability calibration}
The visible energy of the data is corrected for time variation of the mean gain and detector response using muon-induced spallation neutrons captured on Gd and H.
In addition, $\alpha$ decays of $^{212}$Po are collected using the  $^{212}$Bi- $^{212}$Po decay chain coincidence signals to monitor stability.
The energy of the $\alpha$ is 8.8\,MeV but results in a visible energy of about 1\,MeV due to quenching effects, and therefore allows stability monitoring in the lower energy region below neutron capture peaks.

The mean gain of all channels is determined in order to equalize the total number of photoelectrons calculated from the observed charge to the expectation from the hit PMT multiplicity ($n$).
Time variation of the mean gain is measured to be 1.2\,\% from the standard deviation of the H capture peak of muon-induced spallation neutrons collected during the physics data taking.
Because the magnitude of the variation is energy dependent due to the residual gain non-linearity and inefficiency of single p.e. detection, the time variation correction is applied with a linear dependence on energy.
The energy dependence is determined using $\alpha$ decays of $^{212}$Po and neutrons captured on H and Gd by minimizing the time variation of the peak energies for the three samples.
After the gain calibration is applied, the remaining time variation is considered to be due to the detector response, such as scintillator light yield and readout response.
The time variation of the detector response is measured using the peak energy of neutrons captured on H distributed over the NT and GC to be $+0.30$\,\%/year, increasing with time, which is calibrated out as part of stability correction.
The exact source of the increase is under investigation.

\begin{figure}
\begin{center}
	\includegraphics[width=8cm]{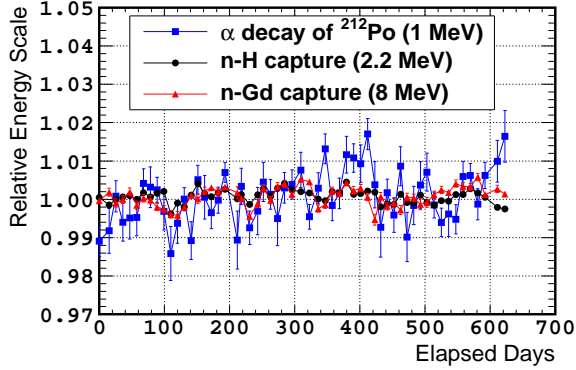}
	\caption{Ratio of the peak energy of $^{212}$Po $\alpha$ decay (blue) and neutron captures on H (red) and Gd (black), after application of the stability correction, to their nominal value as a function of time. Error bars show the statistical uncertainty of the peak energy.}
	\label{fig:PeakStability}
\end{center}
\end{figure}

Figure~\ref{fig:PeakStability} shows the stability of the peak energy of $^{212}$Po $\alpha$ decays and neutron captures on H and Gd after the stability correction is applied.
Time variations of the visible energy are measured to be 0.70\,\% at 1\,MeV, 0.17\,\% at 2.2\,MeV and 0.25\,\% at 8\,MeV from the standard deviations of the peak energies.
The H capture peak is the most stable among the three samples because it is used to extract the time variation.
The systematic uncertainty of the stability is evaluated to be 0.50\,\% from the remaining time variation after correction using $\alpha$ decays of $^{212}$Po and neutron captures on Gd weighted to the IBD energy spectrum.

\subsubsection{Energy non-linearity calibration}
The visible energy of the MC is corrected for its non-linearity relative to that of the data, which arises from two different sources: charge non-linearity (QNL) and light non-linearity (LNL).
The correction function is given as follows:
\begin{eqnarray}
f_{\rm nl}(E_{\rm vis}^{0}) &=& (0.0023 \times E_{\rm vis}^{0}{\rm [MeV]} + 0.9949)
                                                \times (-0.027/E_{\rm vis}^{0}{\rm [MeV]} + 1.008).
\label{eq:NonLinearity}
\end{eqnarray}
The first factor represents the QNL correction, which is associated with the modeling of the readout system and charge integration algorithm and therefore applied to the visible energy of all events, whereas the LNL correction, the second factor, is applied only to the prompt (positron) signals as it arises from the scintillator modeling, which is particle dependent.

The QNL correction is determined using the calibration data with $^{252}$Cf neutron source at the center of the detector.
Although the total energy of $\gamma$'s from neutron capture on Gd is 8\,MeV, the average energy of each $\gamma$ is about 2.2\,MeV which is almost the same as that from neutron capture on H.
Therefore, the discrepancy of the energy response between the data and MC in the ratio of the Gd and H capture peaks can be understood as a consequence of systematic bias in the modeling of readout and the charge reconstruction algorithm.

\begin{figure}
\begin{center}
	\includegraphics[width=8cm]{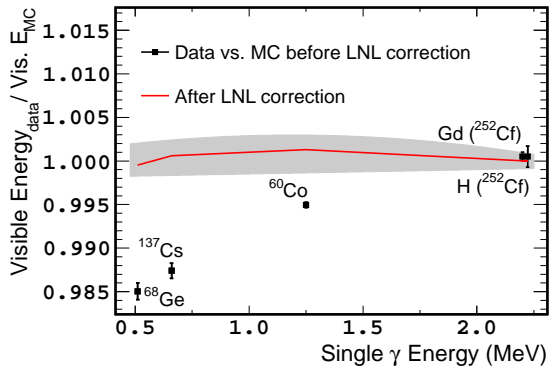}
	\caption{Points show the remaining discrepancy of the energy scales between the data and MC after QNL correction is applied. Horizontal axis shows the average single $\gamma$ energy for various calibration sources deployed at the center of the NT. The $\gamma$ multiplicity of Gd captures has been evaluated using MC. Red line shows the agreement after LNL correction is applied with the systematic uncertainty shown by the grey band.}
	\label{fig:LNL}
\end{center}
\end{figure}

Figure~\ref{fig:LNL} shows the remaining discrepancy of the energy scales between the data and MC after the QNL correction is applied.
The plot is shown as a function of the average single $\gamma$ energy for various $\gamma$ and $^{252}$Cf neutron sources deployed at the center of the NT.
The remaining discrepancy shows a dependence on single $\gamma$ energy, not the total visible energy.
This indicates that the cause of the discrepancy is not in the charge reconstruction but in the scintillator modeling.
In order to evaluate the LNL, MC simulations are generated with several different combinations of Birks' quenching parameter $kB$ and the light yield of the liquid scintillator within the uncertainties from the lab measurements~\cite{ref:Birks, ref:LY}.
Varying the light yield of the scintillator changes the ratio of scintillation light to Cherenkov light and results in different non-linearity. 
Among the several combinations, possible sets of the two parameters, those which show reasonable agreement with data, are retained and a positron MC simulation is generated with each of these sets of parameters.
From a comparison of those positron MC simulations with the one produced with the nominal set of the two parameters, the correction function and the systematic uncertainty are evaluated.

The systematic uncertainties on the non-uniformity, instability, QNL and LNL corrections are collectively represented by a second-order polynomial as a function of $E_{\rm vis}$ (see Section~\ref{section:RS}).
The spread in the number of IBD candidates resulting from varying the 4\,MeV low energy cut on the delayed energy within these systematic uncertainties is $<$0.1\,\%.

\begin{figure}
\begin{center}
	\includegraphics[width=11cm]{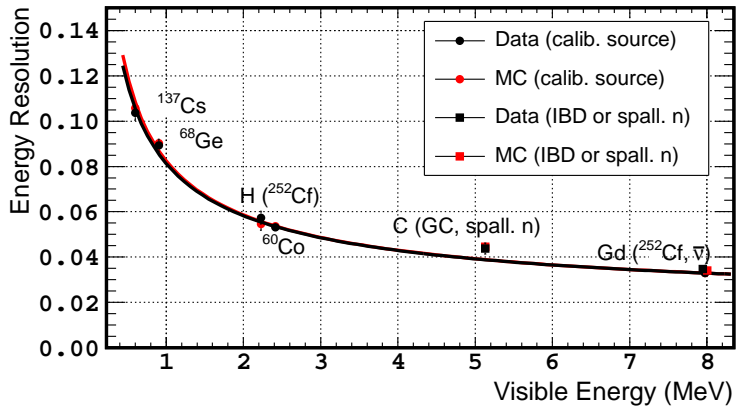}
	\caption{Comparison of the visible energy of the data and MC simulation. Horizontal axis shows the peak energy obtained by a fit and vertical axis shows the energy resolution. Black circles show the data taken with calibration sources at the center of NT and black squares show the peak energy and resolution of neutron captures on Gd in the NT and captures on C distributed over the GC. Red circles and squares show those from the MC simulation. The lines show the fit function of the resolution described in the text.}
	\label{fig:FracWidth}
\end{center}
\end{figure}

Figure~\ref{fig:FracWidth} shows a comparison of the visible energy of the data and MC simulation and their resolution at various energies.
Both the energy scales and the energy resolution show good agreement.
The energy resolution is fit with a function~\cite{ref:PDG}: $\sigma/E_{\rm vis} = \sqrt{a^{2}/E_{\rm vis} + b^{2} + c^{2}/E_{\rm vis}^{2}}$, where $a$, $b$ and $c$ represent the statistical fluctuation, constant term and energy independent width due to electronic noise, respectively.
The best fit to the data is given with $a=0.077 \pm 0.002\,{\rm MeV^{1/2}}$, $b=0.018 \pm 0.001$ and $c=0.017 \pm 0.011\,{\rm MeV}$ and the best-fit to the MC gives  $a=0.077 \pm 0.002\,{\rm MeV^{1/2}}$, $b=0.018 \pm 0.001$ and $c=0.024 \pm 0.006\,{\rm MeV}$.
All parameters are consistent between the data and MC.
Systematic uncertainties on energy scale are summarized in Table~\ref{table:EnergyUncertainty}. 

\begin{table}[ht]
  \footnotesize
  \begin{center}
    \begin{tabular}{| l | c | c |}
      \hline
      Source & Uncertainty (\%) & Gd-III/Gd-II\\
      \hline
      Non-uniformity & 0.36 & 0.84 \\
      Instability & 0.50 & 0.82 \\
      Non-linearity & 0.35 & 0.41\\
      \hline
      Total & 0.74 & 0.65 \\
      \hline
    \end{tabular}
  \end{center}
  \caption{Systematic uncertainties on energy scale. Uncertainty due to non-linearity and the total uncertainty are independently calculated as weighted averages of the prompt energy spectrum, and therefore, the total uncertainty is not equal to the quadratic sum of each uncertainty in this table. Gd-III/Gd-II represents the reduction of uncertainties with respect to the previous publication~\cite{ref:DCII_nGd}.}
  \label{table:EnergyUncertainty}
\end{table}

\subsection{Muon Track Reconstruction}
\label{section:MuonTrackReconstruction}
Cosmic muons passing through the detector often create cosmogenic isotopes in the NT by spallation interactions.
Among these, $\beta$-$n$ emitters, such as $^{9}$Li and $^{8}$He, are the dominant background source in Double Chooz.
The $\beta$-$n$ emitters are hardly distinguished from the IBD signal using the event topology, while spatial correlations with the preceding energetic muons are used to preferentially identify the candidates.
Two muon tracking algorithms are used in the analysis.
One utilizes the spatial pattern of the PMT hit times in the ID, which was used in the previous publications, and also used this time to evaluate the $^{9}$Li and $^{8}$He background rates (see Section~\ref{section:CosmogenicBG}).
In addition, a new algorithm has been developed using full detector information, that is the ID, IV and OV (details are found in Ref.~\cite{ref:FIDO}).
This new algorithm is employed in the calculation of the $^{9}$Li likelihood, which is used to suppress the background (see Section~\ref{section:BGreduction}) and also to measure the $^{9}$Li+$^{8}$He energy spectrum.

\section{Neutrino Selection}
\label{section:Selection}
In Double Chooz, events are recorded at a trigger rate exceeding 150\,Hz while the expected reactor $\bar\nu_{e}$ signal rate is approximately 50\,events/day.
Therefore, effective event selection is required to select IBD signal from the large amount of background.
Background from random singles is largely suppressed by requiring time and space correlations between the two triggers.
Correlated backgrounds, mostly induced by cosmic muons, is suppressed using their characteristic features.
Stopping muons are mostly identified by the IV and OV, but often enter from the chimney where the IV is not sensitive.
Such events, with the vertex position inside the chimney, have a different hit PMT pattern from IBD signal in the NT due to different acceptance.
Fast neutrons often leave energy deposits in the IV and can be distinguished from IBD signal.
Cosmogenic isotopes, such as $^{9}$Li and $^{8}$He, have the same event topology as the IBD signal but can be suppressed using the correlations with the parent muons.
More details of each background are described in Section~\ref{section:Background}, while the selection criteria for the IBD signal are explained in this section.

\subsection{Single Event Selection}
\label{section:SingleSelection}
As the first step of neutrino selection, primary cuts are applied.
Events with a visible energy below 0.4\,MeV are rejected.
An event is tagged as a muon and rejected if it satisfies $E_{\rm vis} > 20\,{\rm MeV}$ or  $E_{\rm IV} > 16\,{\rm MeV}$, where $E_{\rm IV}$ is the energy deposited in the IV.
In addition, events following a muon within a 1\,ms time window are also rejected as these events are mostly background induced by spallation neutrons and cosmogenic isotopes.
Inefficiency due to muon veto is 4.49\,\% with uncertainty of $<$0.01\,\%.

\begin{figure}
\begin{center}
	\includegraphics[width=8cm]{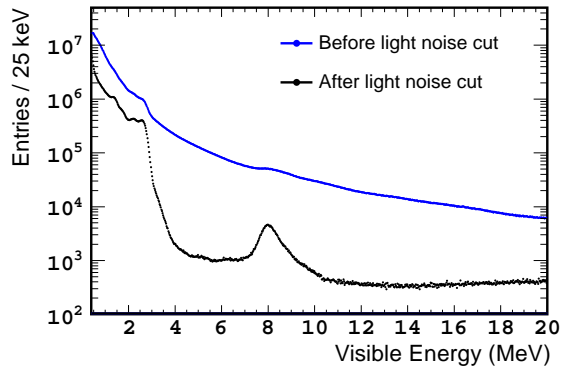}
	\caption{Visible energy of the data. Blue and black histograms show the energy spectrum before and after the light noise cuts are applied, respectively.}
	\label{fig:LightNoise}
\end{center}
\end{figure}

Events which satisfy at least one of the following criteria are discarded as {\it light noise}, a known background caused by a spontaneous light emission from some PMT bases: 
1)  $q_{\rm max}/q_{\rm tot} > 0.12$, where $q_{\rm max}$ and $q_{\rm tot}$ are the maximum charge recorded by a PMT and the total charge in the event, respectively; 
2) $\sigma_t >$ 36\,ns and  $\sigma_q > (464 - 8 \sigma_t$)\,CU (charge unit), where $\sigma_t$ and $\sigma_q$ are the standard deviation of the PMTs hit time and integrated charge distributions, respectively; 
3) $Q_{\rm dev} >  3 \times 10^{4}$\,CU where $Q_{\rm dev}$ is defined as: $Q_{\rm dev} = 1/N \times \sum_{i}^{N}{(q_{\rm max} - q_{i})^{2}/q_{i}}$, where $N$ is the number of PMTs within a sphere of 1\,m radius centered at the PMT with maximum charge.
The first criterion was used in the previous publication but the cut condition is relaxed to minimize inefficiency of the IBD signals at low energy.
Criteria 2) and 3) are new cuts introduced for this analysis.
A one-dimensional cut on $\sigma_t$ in the previous analysis is replaced by a two-dimensional cut on $\sigma_t$ and $\sigma_q$, resulting in an improved rejection performance despite the looser cut on $\sigma_t$ for low energy events, which enabled the reduction of the analysis threshold.
The $Q_{\rm dev}$ calculation is a measure of the non-uniformity of the observed charge of the neighboring PMTs around the one with maximum charge, which tends to be large for light noise events.
The $Q_{\rm dev}$ cut is effective in reducing light noise at high energy around or above the Gd capture peak.
Figure~\ref{fig:LightNoise} shows the reduction of light noise background by the above cuts.
The Gd capture peak and the spectra of radioactive contaminants, such as the 2.6\,MeV $\gamma$ from $^{208}$Tl, are visible after light noise events are rejected.
The inefficiency for IBD signals due to light noise cuts is estimated to be $0.0124 \pm 0.0008$\,\% by the MC simulation.
The remaining light noise is further reduced by a cut on the event likelihood as explained in Section~\ref{section:BGreduction}.

\subsection{IBD Selection}
\label{section:IBDSelection}
After the primary cuts are applied, IBD candidates are selected by the requirement of a delayed coincidence.
This requirement consists of the following selections: 
1) visible energy of the prompt signal should satisfy $0.5 < E_{\rm vis} < 20\,$MeV; 
2) the delayed signal should satisfy $4 < E_{\rm vis} < 10\,$MeV; 
3) the time correlation between the prompt and delayed signals should be in the $0.5 < \Delta T < 150\,\mu$s time window; 
4) the distance between the vertex positions of the prompt and delayed signals should be within $\Delta R < 100\,$cm; 
5) there are no additional events except for the delayed signal found within 200\,$\mu$s before and 600\,$\mu$s after the prompt signal (multiplicity cut).

\begin{figure}
\begin{center}
	\includegraphics[width=8cm]{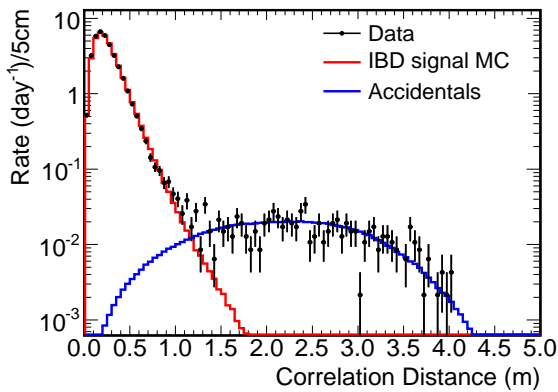}
	\caption{Distance between the vertex positions of the prompt and delayed signals. Points show the data and red line shows the IBD signal MC. Blue line shows the data collected in off-time coincidence windows.}
	\label{fig:dR}
\end{center}
\end{figure}

Signal windows are enlarged with respect to the previous analysis.
The prompt energy window is extended in both directions  to give stronger constraints on the background in the final fit.
The delayed energy window is shifted to lower energies to improve the IBD signal efficiency and reduce the systematic uncertainty on the neutron detection efficiency.
The cross-section of neutron capture on Gd has a strong energy dependence and therefore the probability distribution of neutron capture in the MC simulation is sensitive to the neutron scattering model for a few microseconds after the IBD reaction.
The lower cut on $\Delta T$ was set at 2\,$\mu$s in the previous analysis due to contamination of stopping muons below the cut, thus introducing a systematic uncertainty.
This is reduced in the new analysis in which stopping muons are largely suppressed using new vetoes (see Section~\ref{section:BGreduction}) and, as a consequence, the $\Delta T$ signal window is extended to lower times to contain neutrons in the thermalization phase.
In the previous publications, the $\Delta R$ cut was applied only to select neutron captures on H in which accidental background is dominant, whereas this cut is now applied in the new analysis to select Gd captures.
Figure~\ref{fig:dR} shows the $\Delta R$ distributions for the data and MC simulation, together with the data collected in off-time coincidence windows (see Section~\ref{section:AccidentalBG}).
The accidental background is suppressed by the $\Delta R$ cut, while the IBD signal inefficiency is 0.3\% which is included in the selection efficiency (see Section~\ref{section:IBDefficiency}).
The reduction of accidental background enables us to extend the delayed energy window down to 4\,MeV.

Inefficiency due to multiplicity cut is precisely measured by the single event rate as 1.06\% with $<$0.01\,\% uncertainty.
Efficiency of the IBD selection besides the multiplicity cut is evaluated to be $98.4$\,\% using the signal MC, where efficiency is defined as the ratio of the number of events selected with the IBD selection to that selected by the extended signal window: $3.5 < E_{\rm vis} < 12\,$MeV for delayed signal;  $0.25 < \Delta T < 1000\,\mu$s; and no cut on $\Delta R$.
With the same definition, the efficiency was 91.2\,\% by the selection criteria used in the previous publication~\cite{ref:DCII_nGd}.

\subsection{Background Reduction}
\label{section:BGreduction}
After the delayed coincidence is required, the remaining backgrounds are mostly induced by cosmic muons: i.e. long-lived cosmogenic isotopes, proton recoils by spallation neutrons (referred to as fast neutrons) and stopping muons.
In order to suppress such backgrounds, the following vetoes are applied.

\begin{figure}
\begin{center}
	\includegraphics[width=8cm]{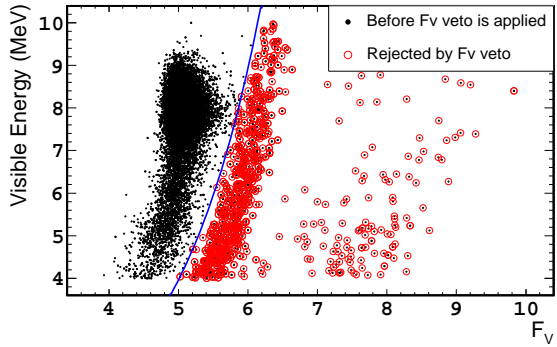}
	\caption{Correlations between $F_{\rm V}$ and visible energy for the delayed signals. Black points show the data before {\it $F_{\rm V}$\,veto} is applied, and red circles on top of the black points mark the events rejected by {\it $F_{\rm V}$\,veto}.  The blue line shows the selection criterion of  {\it $F_{\rm V}$\,veto}.}
	\label{fig:FVveto}
\end{center}
\end{figure}

\paragraph{\it $F_{\rm V}$\,veto:} 
$F_{\rm V}$ is given by Eq.~\ref{eq:FuncV}.
$F_{\rm V}$ becomes large for events which have a different hit pattern than a point-like source in the NT and GC, such as stopping muons which enter through and stop inside the chimney, where IV and lower OV are not sensitive, and light noise events.
The delayed signal should satisfy $E_{\rm vis} > 0.068 \times \exp{(F_{\rm V}/1.23)}$.
Figure~\ref{fig:FVveto} shows the correlations between $F_{\rm V}$ and visible energy for the delayed signals, where we can find three components: lower $F_{\rm V}$ band by IBD signals, middle band by stopping muons and events with higher $F_{\rm V}$ due to light noise not rejected by the cuts described in Section~\ref{section:SingleSelection}.
{\it $F_{\rm V}$\,veto} effectively rejects these background events.

\paragraph{\it OV veto:} 
Stopping muons are also excluded by {\it OV veto}.
If the prompt signal is coincident with OV hits within 224\,ns, the event is rejected.

\paragraph{\it IV veto:} 
{\it IV veto} is motivated to reduce fast neutron background events which often make recoil protons and deposit energy in the IV, below the threshold of muon identification.
If the prompt signal satisfies all the following conditions, the event is rejected: IV PMT hit multiplicity $\ge$ 2; total integrated charge in the IV $>$ 400\,CU (corresponding to roughly 0.2\,MeV); outputs of the event reconstruction in the ID and IV are close in space ($< 3.7\,$m) and time (within $50\,$ns).

\paragraph{\it Li+He veto:} A $^{9}$Li likelihood is calculated for each combination of prompt event and preceding muon based on: the distance between the event vertex position to the muon track and the number of neutron candidates following the muon within 1\,ms.
Probability density functions (PDF) for each variable are produced from muon-induced $^{12}$B collected during data taking instead of $^{9}$Li events to accumulate statistics.
It is confirmed that the PDFs from $^{12}$B agree with those from $^{9}$Li.
Prompt signals which satisfy $\mathcal{L}_{\rm Li} $ cut condition are rejected as $^{9}$Li or $^{8}$He candidates, where $\mathcal{L}_{\rm Li}$ is the maximum $^{9}$Li likelihood among all combinations with the preceding muons within 700\,ms.
The {\it Li+He veto} rejects $1.12\pm0.05$\,events/day, which corresponds to 55\,\% of the cosmogenic background estimation.

\begin{figure}
\begin{center}
	\includegraphics[width=8.5cm]{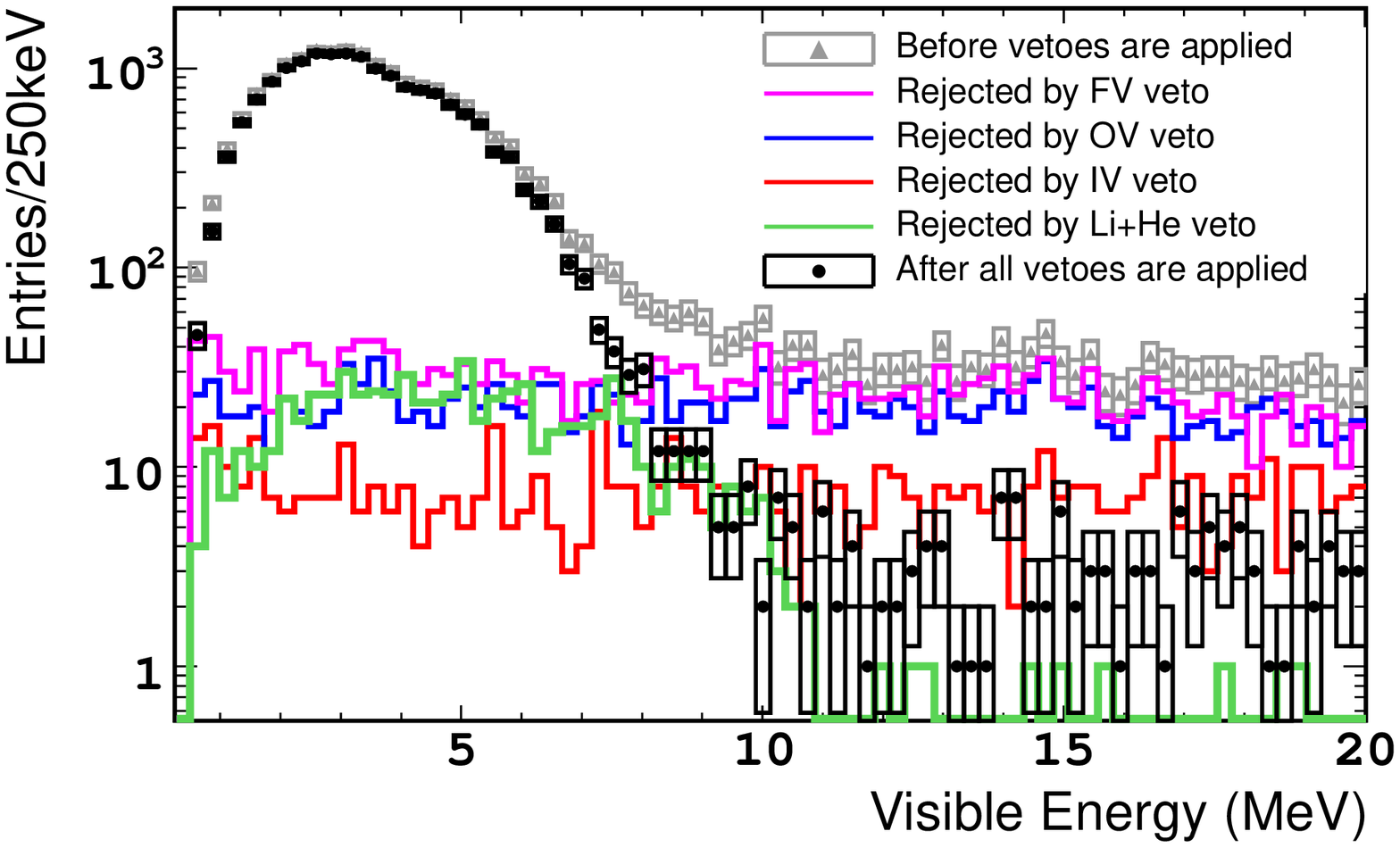}
	\includegraphics[width=8.5cm]{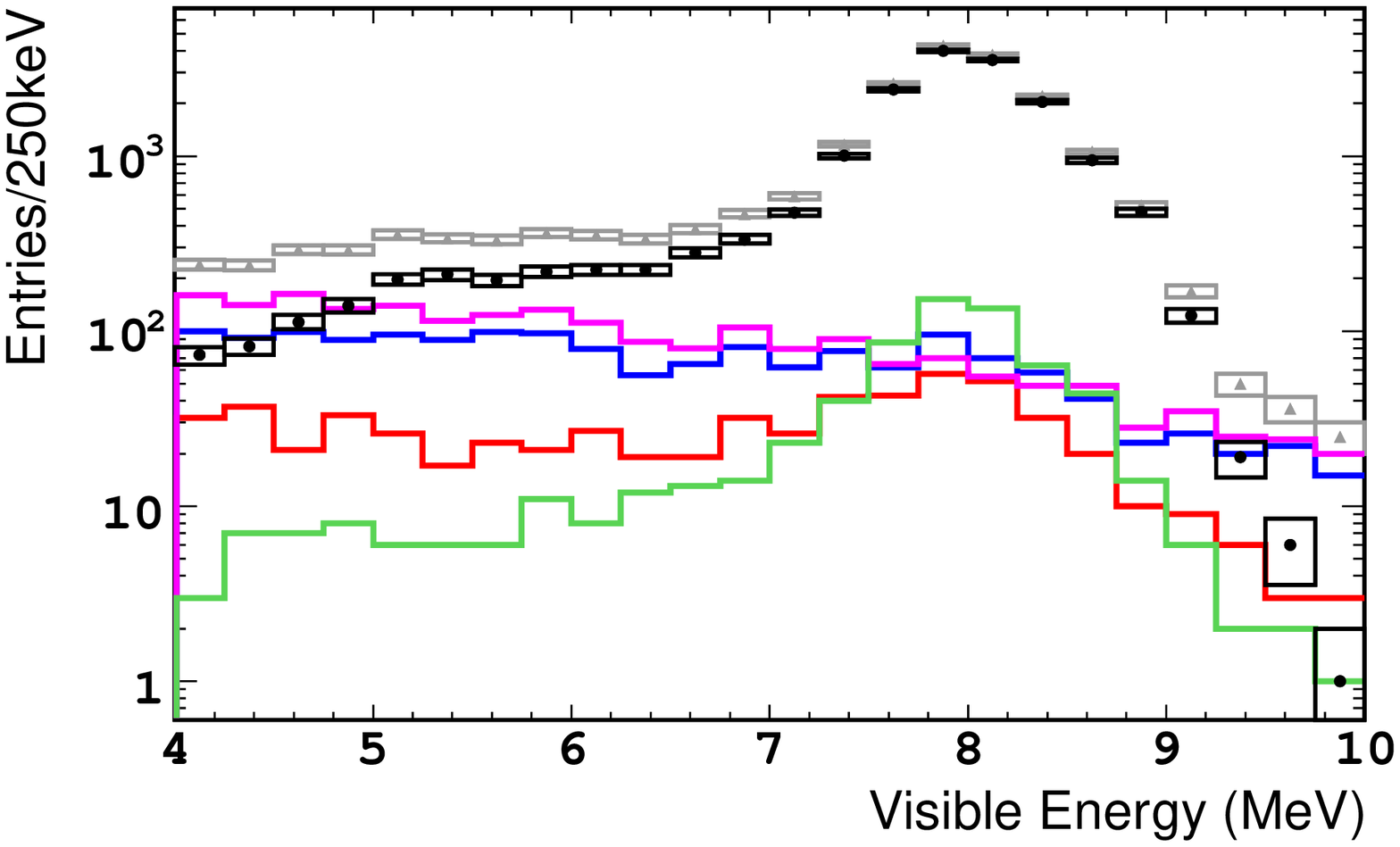}
	\includegraphics[width=8.5cm]{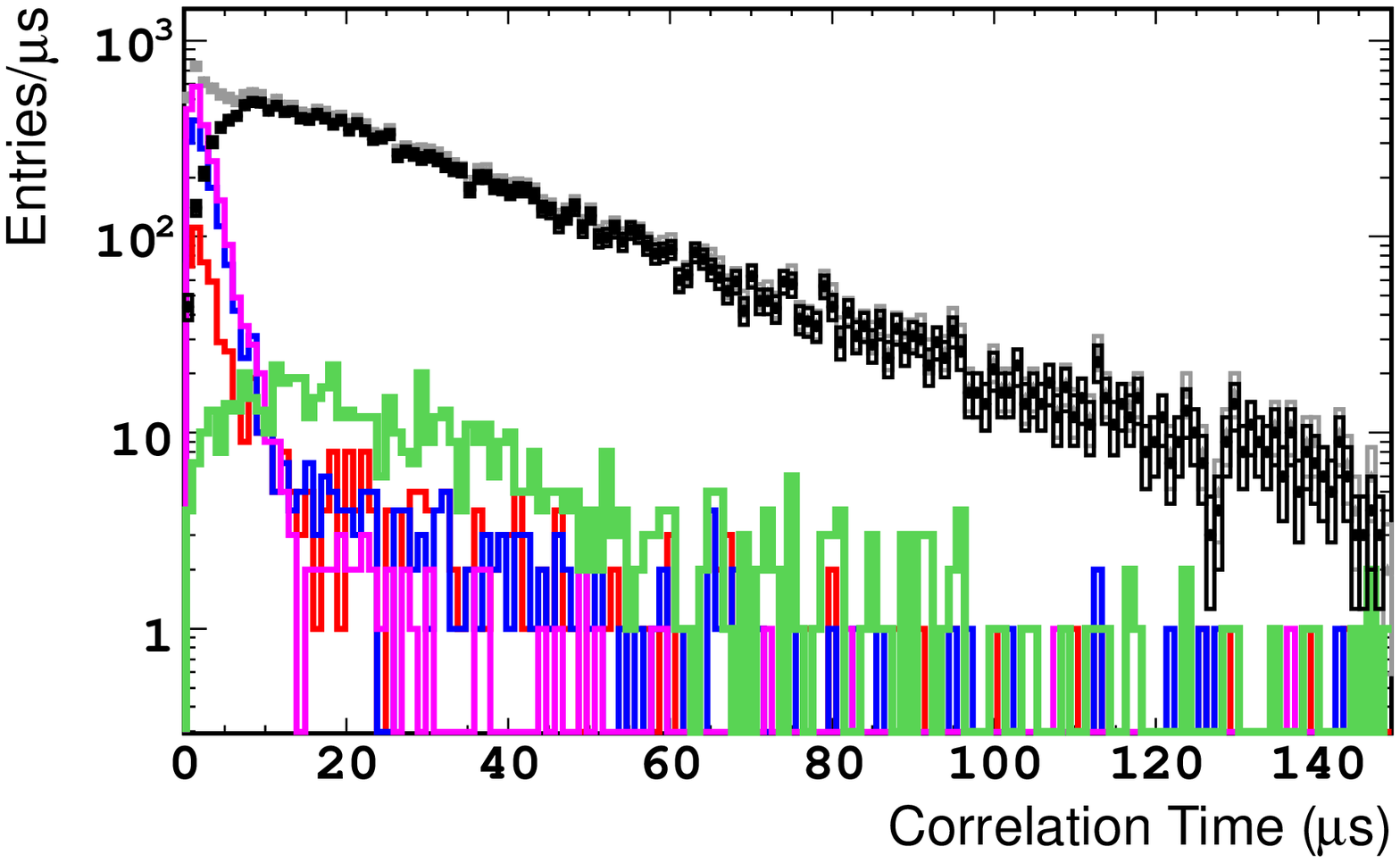}
	\caption{Visible energy of the prompt signals (top) and delayed signals (middle) and correlation time of the prompt and delayed signals (bottom). Points and boxes show the data with their statistical uncertainty before (grey, triangle) and after (black, circle) all analysis vetoes are applied. Lines show the events rejected by each individual veto: {\it $F_{\rm V}$\,veto} (magenta); {\it OV veto} (blue); {\it IV veto} (red); and {\it Li+He veto} (green).}
	\label{fig:vetoes}
\end{center}
\end{figure}

Figure~\ref{fig:vetoes} shows the breakdown of rejected events by each veto and remaining IBD candidates as a function of the visible energy of the prompt and delayed signals and time correlation between them.
Among these, {\it $F_{\rm V}$\,veto} and {\it IV veto} are newly developed for this analysis.
The {\it $F_{\rm V}$\,veto}, {\it OV veto} and {\it IV veto} respectively reject 71\,\%, 62\,\% and 24\,\% of events above 12\,MeV, where fast neutrons and stopping muons are dominant, and a combination of the three vetoes rejects 90\,\% of these high energy events.
The inefficiencies of the IBD signal due to {\it $F_{\rm V}$\,veto}, {\it OV veto} and {\it IV veto} are $0.06\pm0.11\,\%$, $0.058\pm0.001\,\%$ and $0.035\pm0.014\,\%$, respectively.
In the previous analysis, in order to reject $^{9}$Li and $^{8}$He, a longer veto interval was applied after energetic muons resulting in an additional 4.8\,\% dead time.
It is replaced by the likelihood-based cut, for which the inefficiency of IBD signals is only $0.504\pm0.018$\,\% for comparable reduction power.
The energy spectrum and $\Delta T$ of the rejected events are consistent with $^{9}$Li and $^{8}$He as shown in Fig.~\ref{fig:vetoes}.
More details about the background events are described in Section~\ref{section:Background}.

\section{IBD Detection Efficiency}
\label{section:DetectionEfficiency}
Double Chooz is taking data with a dead-time-free data acquisition system and the trigger efficiency reaches 100\,\% at 500\,keV with negligible uncertainty.
The detection efficiency of the prompt signal is determined to be close to 100\,\%.
On the other hand, various physics issues are involved in the detection of the delayed signal, such as Gd concentration, neutron scattering models, energy scale and vertex resolution.
The systematic uncertainty on the detection of the delayed signal is evaluated using calibration data taken with the $^{252}$Cf source and IBD candidate events.
Data are compared with the MC simulation to extract the correction factor for the MC normalization, integrated over the volume, and its systematic uncertainties.
The correction consists of the product of three independent contributions: $C_{\rm GdF}$ corrects for the fraction of neutron captures on Gd; $C_{\rm Eff}$ corrects for the IBD selection efficiency over the full volume; and $C_{\rm Sio}$ corrects for the modeling of spill in/out by the MC simulation.
Each factor and its systematic uncertainty is described in this section.

\subsection{Gd Fraction}
\label{section:GdFrac}
Neutrons in the NT are captured dominantly on Gd and H.
The fraction of neutron captures on Gd depends mainly on the relative Gd concentration in the liquid scintillator.
The fraction is measured using calibration data with a $^{252}$Cf source at the center of the detector as a ratio of the number of coincidence events selected by two different delayed energy windows.
Neutrons are emitted in the spontaneous fission of $^{252}$Cf  together with $\gamma$'s, which are detected as the prompt signal followed by the neutron captures.
The window is set to be $3.5 <  E_{\rm vis} < 10\,{\rm MeV}$ for the numerator to select only Gd capture (small fraction of C captures are also included) and $0.5 < E_{\rm vis} < 10\,{\rm MeV}$ for the denominator to include H captures.
In order to suppress background contamination in the $^{252}$Cf data, two cuts are applied in addition to the standard IBD selection criteria: 
1) $E_{\rm vis} > 4\,{\rm MeV}$ where $E_{\rm vis}$ is the visible energy of the prompt signal; 
2) more than one neutron are detected after the prompt signal ($^{252}$Cf emits 3.8 neutrons on average~\cite{ref:NeutronMultiplicity}).
2) is applied to eliminate a correlated background contribution.
The accidental background is measured using the off-time coincidence windows and subtracted.
The Gd fraction is measured to be $85.30\pm0.08$\,\% for the data and $87.49\pm0.04$\,\% for the MC simulation with the above definition.
The correction factor for the Gd fraction is measured from the ratio of the data to that of the MC to be $C_{\rm GdF} = 0.9750 \pm 0.0011 (\rm stat) \pm 0.0041 (\rm syst)$.
The systematic uncertainty is evaluated by varying the energy windows.
Consistent values are obtained from the $^{252}$Cf calibration data taken at different times and positions, which confirm the Gd concentration in the liquid scintillator is stable and uniform over the NT.
In addition,  the Gd fraction is measured using the IBD candidates and spallation neutrons and they agree within the uncertainties.

\subsection{IBD Selection Efficiency}
\label{section:IBDefficiency}
The correction factor for IBD selection efficiency is evaluated by two methods, one using the $^{252}$Cf calibration data and the other using IBD candidates, both described in this section.
Both methods yield correction factors consistent with unity (i.e. no correction) within the uncertainty of a few per mil.
The final correction factor is determined from the combination of those obtained by the two methods as: $C_{\rm Eff} = 1.0000 \pm 0.0019.$

\subsubsection{Efficiency measurement by Cf neutron source}
\label{sec:VFmethod}
\begin{figure}
\begin{center}
	\includegraphics[width=8.5cm]{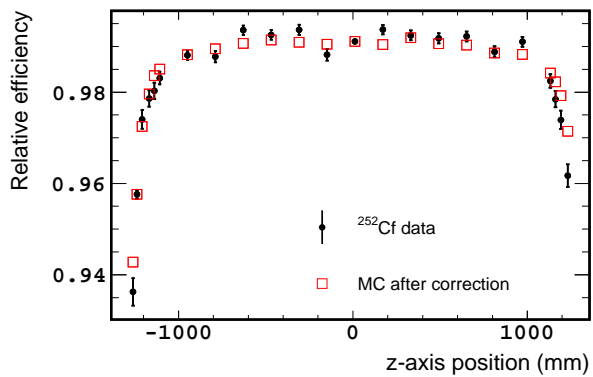}
	\caption{Neutron detection efficiency measured by $^{252}$Cf calibration data along the $z$-axis and the MC simulation. Horizontal axis shows the position where the calibration source is placed. The discrepancy at large $z$ can be explained by the uncertainty on the position at which the source is deployed and the steeply falling efficiency at these values of $z$.}
	\label{fig:VFefficiencyCf}
\end{center}
\end{figure}

The efficiency of the neutron capture signal can be precisely measured using $^{252}$Cf calibration data which have high statistics and negligible background contamination.
The source was deployed along the vertical symmetry axis ($z$-axis) at different levels.
The efficiency is defined as the ratio of the number of events selected by the standard IBD selection to that by the extended signal window: $0.25 < \Delta T < 1000\,\mu{\rm s}$; no $\Delta R$ cut; and $3.5 <  E_{\rm vis} < 10\,{\rm MeV}$ on the delayed signal.
Averaged efficiency over the NT is derived using volume factorization of the efficiency based on the measurements by $^{252}$Cf calibration data.
The efficiency at a certain position $(z, \rho)$ inside the NT is constructed as:
\begin{equation}
\label{eq:VFposition}
\epsilon(z, \rho) = \epsilon_{0} \times f_{1}(z) \times f_{2}(\rho),
\end{equation}
where $\epsilon_{0}$ represents the efficiency at the target center, which is precisely measured by $^{252}$Cf calibration data with high statistics.
$f_{1}(z)$ and $f_{2}(\rho)$ describe the $z$ and $\rho$ dependence at $\rho=0$ and $z=0$, respectively.
$f_{1}(z)$ is measured from $^{252}$Cf calibration data as shown in Fig.~\ref{fig:VFefficiencyCf}, while $f_{2}(\rho)$ is assumed to follow the same behavior as $f_{1}(z)$ as a function of the distance to the NT wall.
Validity of this assumption is confirmed by the MC simulation.
The averaged efficiency over the target volume is then obtained to be $98.29\pm0.23$\,\% for the data and $98.26\pm0.22$\,\% for the MC by integrating Eq.~\ref{eq:VFposition} over the NT volume, in which systematic uncertainties associated with the $f_{1}(z) \rightarrow f_{2}(\rho)$ conversion and integration over the full volume from finite data points are taken into account.
The correction factor for the IBD selection efficiency is obtained by comparing the data and MC as: $C_{\rm Eff, Cf} = 1.0003 \pm 0.0032$, where the error includes both statistical and systematic uncertainties.

\subsubsection{Efficiency measurement by IBD candidates}
\label{section:IBDmethod}
In this method, the correction factor for the efficiency is directly measured using IBD candidates.
Integrated efficiency over the NT volume can be measured using the IBD candidates, while compared to the $^{252}$Cf calibration data the statistics are limited and background contamination is not negligible.
Given these conditions, measurement by the IBD candidates is complementary to that by $^{252}$Cf calibration data.

The efficiency of neutron capture signals is defined as the ratio of the number of IBD candidates selected by the standard signal window to that by the extended one: $0.25 < \Delta T < 200\,\mu{\rm s}$; $\Delta R < 1.7\,{\rm m}$; and $3.5 <  E_{\rm vis} < 10\,{\rm MeV}$ on the delayed signal.
In addition to the standard IBD selection, the following cuts are applied to the prompt signal to reduce background contamination: $0.5 < E_{\rm vis} < 8$\,MeV; and $F_{\rm V} < 5.8$.
Contamination from the accidental background is measured using the off-time coincidence windows and subtracted.

Efficiencies are measured to be $98.58\pm0.19\,\%$ for the data and $98.62\pm0.02\,\%$ for the $\overline{\nu}_{e}$ MC simulation.
Comparing the efficiencies measured by the data and the $\overline{\nu}_{e}$ MC simulation, the correction factor is: $C_{\rm Eff, IBD} = 0.9996 \pm 0.0021$, where the error includes statistical and systematic uncertainties.
The correction factor is also measured using only the IBD candidates whose vertex position is reconstructed at the bottom half of the detector to suppress the contamination from stopping $\mu$ background.
The discrepancy is taken into account as the systematic uncertainty.

The IBD detection efficiency is also measured exclusively for each cut on $\Delta T$, $\Delta R$ and $E_{\rm vis}$ of delayed signals.
The total efficiency is given as a product of them.
The correction factor measured by this exclusive approach is found to be consistent with that written above.
\\

\subsection{Spill-in/out}
\label{section:spill}
Although the IBD selection dominantly collects neutron captures on Gd, it does not ensure that all IBD reactions occur inside the NT. 
Incoming neutrons from IBD reactions outside the NT lead to neutron captures on Gd (spill-in) while neutrons created inside the NT may escape to the GC and be captured on H (spill-out). 
These spill-in and spill-out currents do not cancel each other: there are more spill-in events than spill-out events due to the longer neutron travel distance in the GC in absence of Gd and geometrical considerations. 
The source of the systematic uncertainty on the spill-in and spill-out current is dominated by the modeling of low energy neutrons scattering on light nuclei, and is particularly sensitive to the molecular bonds.
In Double Chooz, a custom neutron simulation code was developed for the neutron thermalization process (see Section~\ref{section:Simulation}).

The net spill current is defined as the difference between the number of spill-in and spill-out events.
As the spill currents cannot be measured from the data directly, the ratio of the net spill current to the number of IBD interactions in the NT is calculated from our nominal Geant4-based IBD simulation MC to be 2.08\,\%.
The uncertainty of this ratio is estimated from the difference between the nominal 2.08\,\% and the alternative value, 2.36\,\%, estimated using a different MC simulation code, Tripoli4~\cite{ref:Tripoli}, designed for the accurate modeling of low energy neutron physics.

Tripoli4 takes molecular bonds into account based on experimental data, whereas the custom neutron scattering model implemented in Geant4 is based on analytical corrections.
Hydrogen atoms are considered to be bonded in CH2 groups, which is the dominant structure in our main scintillator component. Introducing other bonds, as aromatic rings, do not lead to significant discrepancies.
The consistency of Tripoli4 with the data has been studied using Gd fraction and $\Delta T$, highly sensitive observables to the spill current description, demonstrating Tripoli4 to be more accurate than Geant4-based custom model.
The small discrepancies between the data and nominal MC are accommodated by the correction factor for the Gd fraction (see Section~\ref{section:GdFrac}) in the neutrino oscillation analysis.
The uncertainty on spill-in/out effects is obtained from the difference in spill currents by the two simulations to be $C_{\rm Sio} = 1.0000 \pm 0.0027$, which is expected to be conservative.
\\

Total MC normalization correction factor including other sources are summarized in Table~\ref{table:MCCorrection} with their systematic uncertainties.

\begin{table}[ht]
  \footnotesize
  \begin{center}
    \begin{tabular}{| l | c | c |}
      \hline
      Correction source & MC Correction & Uncertainty (\%)\\
      \hline
      DAQ \& Trigger & 1.000 & $<0.1$\\
      Veto for 1\,ms after muon & 0.955 & $<0.1$ \\
      IBD selection & 0.989 & 0.2\\
      FV, OV, IV, Li+He veto & 0.993 & 0.1\\
      Scintillator proton number & 1.000 & 0.3 \\
      Gd fraction & 0.975 &  0.4 \\
      Spill in/out & 1.000 & 0.3\\
      \hline
      Total & 0.915& 0.6 \\
      \hline
    \end{tabular}
  \end{center}
  \caption{Summary of inputs for the MC correction factor and the fractional uncertainties. IBD selection includes correction for inefficiencies due to multiplicity and light noise cuts (see Section~\ref{section:IBDSelection}). Inefficiency due to individual background veto and the uncertainty are shown in Section~\ref{section:BGreduction}.}
  \label{table:MCCorrection}
\end{table}

\section{Backgrounds}
\label{section:Background}
Three types of backgrounds are accounted for in neutrino oscillation analysis: long-lived cosmogenic isotopes, i.e. $^{9}$Li and $^{8}$He resulting in $\beta$-$n$ decays; correlated events due to stopping muons and proton recoils from spallation neutrons; and accidental coincidence of two single events.
In the new analysis, the background rate and energy spectrum shape are estimated by data-driven methods described in this section.

\subsection{Cosmogenic Isotopes}
\label{section:CosmogenicBG}
\begin{figure}
\begin{center}
	\includegraphics[width=8cm]{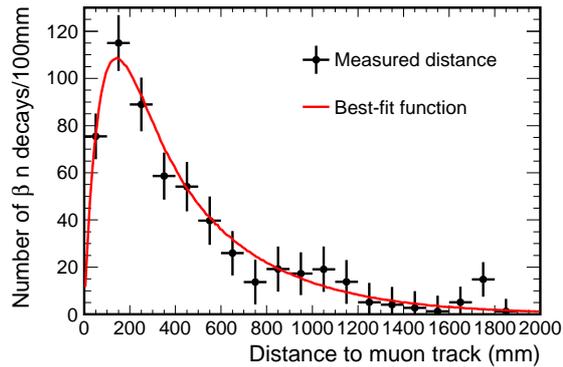}
	\caption{Lateral profile of Li production position with respect to the muon track. Points show the data with the muon energy deposition above 600\,MeV$^{*}$ (explained in the text). Red line shows the best fit of an exponential function ($\lambda_{\rm Li}$ as the mean distance) with a convolution of Gaussian function to account for the resolution of the vertex ($\sigma_{\rm Li})$ and muon track reconstruction ($\sigma_{\mu}$). The fit gives $\lambda_{\rm Li}=(42\pm4)\,$cm and $\sigma_{\mu}=(15\pm4)\,$cm ($\sigma_{\rm Li}=10\,$cm is fixed in the fit).}
	\label{fig:LiProfile}
\end{center}
\end{figure}

Radioisotopes are often produced in spallation interactions of cosmic muons inside the detector.
Some of the cosmogenic isotopes, such as $^{9}$Li and $^{8}$He, emit a neutron in association with their $\beta$ decay, and cannot be distinguished from the IBD signals by the event topology.
The lifetime of $^{9}$Li and $^{8}$He are 257\,ms  and 172\,ms, respectively, much longer than the 1\,ms muon veto.
Contamination from the cosmogenic isotopes (collectively referred to as Li hereafter as the main contribution is $^{9}$Li) is evaluated from fits to the time correlation between the IBD candidates and the previous muons ($\Delta T_{\mu}$). 
The Li rate is evaluated without the {\it Li+He veto} (see Section~\ref{section:BGreduction}) first, and then the fraction of vetoed events is subtracted later.
Muons are divided into sub-samples by the energy in the ID, as the probability to generate cosmogenic isotopes increases with the energy deposits in the detector.
Only the sample above 600\,MeV$^{*}$ (MeV$^{*}$ represents MeV-equivalent scale as the energy reconstruction is not ensured at such high energy due to non-linearity associated with flash-ADC saturation effects) is sufficiently pure to produce a precise fit result. 
At lower energies below 600\,MeV$^{*}$, an additional cut on the distance of muon tracks to the vertex of the prompt signal ($d$) is introduced to reduce accidental muon-IBD pairs: only muons which satisfy $d<75\,{\rm cm}$ are considered.
The inefficiency of muon-Li pairs due to the distance cut is evaluated for each energy range as the product of the acceptance and the lateral profile of the Li vertex position with respect to the muon track.
The lateral profile is extracted from the high energy muon sample above 600\,MeV$^{*}$ as shown in Fig.~\ref{fig:LiProfile}.
After the correction for inefficiency, the total cosmogenic background rate is determined to be $2.20^{+0.35}_{-0.27}$ events/day.

\begin{figure}
\begin{center}
	\includegraphics[width=8cm]{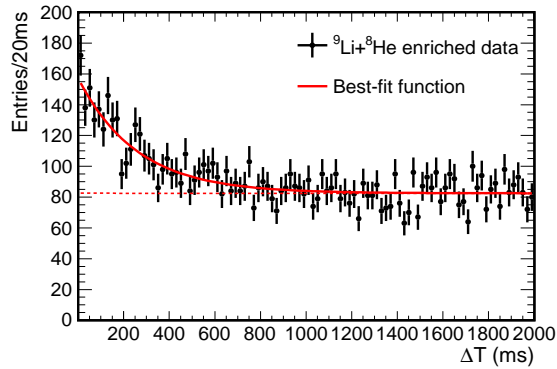}
	\caption{$\Delta T_{\mu}$ distribution of the Li enriched sample. The red line shows the best fit to an accidental coincidence of muons (flat, dashed line) and Li contribution (exponential, solid curve).}
	\label{fig:LiDtMuon}
\end{center}
\end{figure}

In order to further constrain the background rate, a lower limit is computed separately.
A Li-enriched muon sample is selected with the following cuts: 1) $E_{\mu} > 300\,$MeV$^{*}$ if there is more than or equal to one neutron candidates following the muon within 1\,ms; 2) $E_{\mu} > 500\,$MeV$^{*}$ and $d < 0.75\,$m if there is no neutron candidate.
Figure~\ref{fig:LiDtMuon} shows the $\Delta T_{\mu}$ distribution of the Li enriched sample.
The energy cuts are optimized to select a maximum amount of Li candidates while at the same time keeping the accidental muon-IBD pairs as low as possible to minimize uncertainty on the fit parameter. 
The component from cosmogenic isotopes background in the Li enriched sample is found to be $2.05\pm0.13$ events/day from a fit to the $\Delta T_{\mu}$ distribution.
This value is used to set the lower limit.

The rate estimates are combined, yielding a cosmogenic background rate of $2.08^{+0.41}_{-0.15}$ events/day.
The error includes the systematic uncertainties evaluated by varying the cuts on $d$, values of $\lambda_{\rm Li}$ and binning of $\Delta T_{\mu}$ distribution.
In addition, the impact of $^{8}$He is also evaluated assuming a fraction of $8 \pm 7\,\%$ based on the measurement by KamLAND~\cite{ref:KamLAND_He}, rescaled to account for the different energies of the cosmic muons illuminating the two experiments, and taken into account in the rate estimate and its uncertainty.

In the standard IBD selection, Li candidates are rejected by the {\it Li+He veto}.
The number of Li events rejected by the {\it Li+He veto} is determined by a fit to the $\Delta T_{\mu}$ distribution to be $1.12\pm0.05$ events/day.
A consistent value is confirmed by a counting approach, in which the number of Li candidates in the off-time windows is subtracted from the number of Li candidates rejected in the IBD selection.
After subtracting Li events rejected by the {\it Li+He veto}, the final cosmogenic isotope background rate is estimated to be $0.97^{+0.41}_{-0.16}$ events/day.

\begin{figure}
\begin{center}
	\includegraphics[width=8cm]{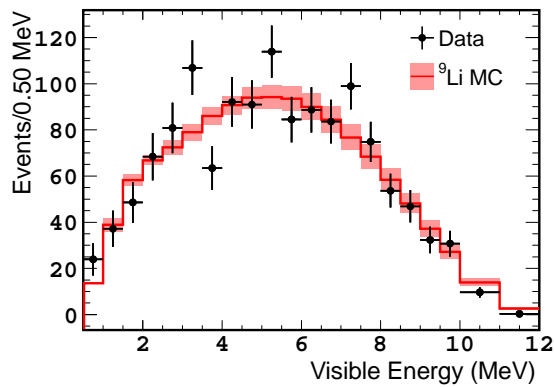}
	\caption{Prompt energy spectrum of cosmogenic background measured by Li candidates. Points show the data with their statistical uncertainties. Overlaid histogram and the band show the prediction from the MC simulation, which includes only Li, and its uncertainty. The MC is normalized to the data.}
	\label{fig:LiMeasuredSpectrum}
\end{center}
\end{figure}

The spectrum shape of cosmogenic isotope background is measured by the Li candidate events which include both $^{9}$Li and $^{8}$He events.
Li candidates with neutrons captured on H are also included to reduce statistical uncertainty.
Backgrounds in the Li candidates (which are due to accidental pairs of muons and IBD signals) are measured by off-time windows and subtracted.
The measured prompt energy spectrum is shown in Fig.~\ref{fig:LiMeasuredSpectrum}, together with the prediction from the $^{9}$Li MC simulation, as reference, which has been newly developed by considering possible branches of the $\beta$-decay chains including $\alpha$ and neutron emissions.

\subsection{Fast Neutrons and Stopping Muons}
\label{sec:FN-SM}
Fast neutrons, induced by spallation interactions of muons in the rock near the detector, can penetrate the detector and interact in the NT or GC, producing recoil protons.
Such events can be background if the recoil protons are detected in the prompt energy window and, later, a thermalized neutron (either the same neutron or a different one) is captured on Gd.
In addition, if a cosmic muon entering the ID through the chimney stops inside the detector and produces a Michel-electron from its decay, the consecutive triggers by the muon and the electron can be a background.
Fast neutrons and stopping muons are collectively referred to as correlated background and the total background rate and energy spectrum shape are estimated.
Contributions from the fast neutrons and stopping muons were comparable in the previous analysis, whereas with the {\it $F_{\rm V}$ veto} introduced in the new analysis, stopping muons are largely suppressed and the remaining background is mostly from fast neutrons.

\begin{figure}
\begin{center}
	\includegraphics[width=9cm]{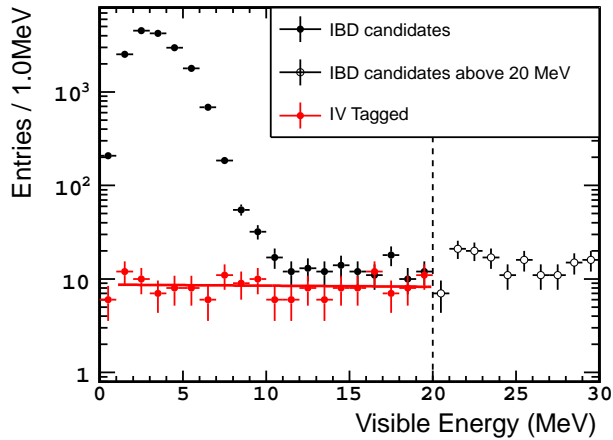}
	\caption{Prompt energy spectrum of three data samples: IBD candidates (black filled points); IV tagged events (red points); and coincident signals above 20\,MeV (black empty circles). The red line shows the best fit of a linear function to the IV tagged events with a slope of  $- 0.02 \pm 0.11\,{\rm events/MeV^{2}}$. IV-tagged events below 1\,MeV are not used in the fit to avoid contamination from Compton scattering of $\gamma$'s in the IV and ID.}
	\label{fig:IVtagged}
\end{center}
\end{figure}

The background spectrum shape is measured using events, referred to as IV-tagged events, which pass the IBD selections except for the  {\it IV veto} but would have been rejected by the {\it IV veto}.
As the fast neutrons and stopping muons often deposit energy in the IV, IV tagging favorably selects correlated background events.
Figure~\ref{fig:IVtagged} shows the prompt energy spectrum of three samples: 1) IBD candidates; 2) IV-tagged events; and 3) coincidence signals above 20\,MeV which are selected by the standard IBD selection but for which the muon veto condition is changed from 20\,MeV to 30\,MeV.  
A slope of $-0.02 \pm 0.11\,{\rm events/MeV^{2}}$ is obtained from a fit to the IV-tagged events with a linear function, which is consistent with a flat spectrum and no evidence for an energy-dependent shape.
The flat spectrum shape is also confirmed with OV vetoed events, and it is consistent with the IBD candidates above 12\,MeV as well, where the correlated background is dominant.
Given these observations, a flat spectrum shape of correlated background is adopted in the neutrino oscillation fit using the energy spectrum.

The correlated background rate is estimated to be $0.604 \pm 0.051$\,events/day from the number of coincident signals in the energy window between 20 and 30\,MeV shown in Fig.~\ref{fig:IVtagged}. 
For the reactor-off running (see Section~\ref{section:ReactorOff}), the background rate is slightly different due to different configurations of the OV from the whole period  (see Section~\ref{section:Detector}), and it is estimated to be $0.529 \pm 0.089$\,events/day.

\subsection{Accidental Background}
\label{section:AccidentalBG}

\begin{figure}
\begin{center}
	\includegraphics[width=8cm]{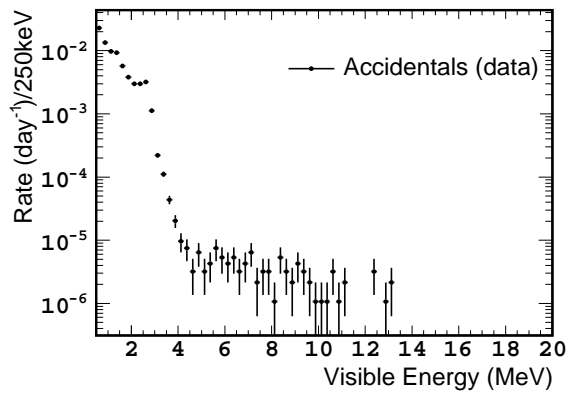}
	\caption{The prompt energy spectrum of the accidental background measured by the data collected using off-time windows.}
	\label{fig:AccidentalBGSpectrum}
\end{center}
\end{figure}

Random associations of two triggers which satisfy the IBD selection criteria are referred to as accidental background.
The accidental background rate and spectrum shape are measured by the off-time window method, in which the time windows are placed more than 1 sec after the prompt candidate, keeping all other criteria unchanged, in order to collect random coincidences only.
A multiple number of successive windows are opened to accumulate statistics.
The background rate in the off-time windows is measured to be $0.0701 \pm 0.0003 (\rm stat) \pm 0.0026 (\rm syst)$ events/day, in which corrections for the different dead time from the standard IBD selection and the associated systematic uncertainties on the correction are accounted for.
The error on the accidental background rate estimate is larger than that in the previous analysis due to a correction factor introduced to account for the different efficiency of the {\it Li+He veto} for accidental coincidences in on-time and off-time windows.
The accidental background rate is found to be stable over the data taking period.
The prompt energy spectrum of the measured accidental background is shown in Fig.~\ref{fig:AccidentalBGSpectrum}.
\\

Estimated background rates are summarized in Table~\ref{table:BGSummary} including contributions from other background sources not used in the neutrino oscillation fit.
The background rate of $^{13}$C($\alpha$, $n$)$^{16}$O reactions is evaluated from the contamination of $\alpha$ emitters (including $^{152}$Gd) in the detector to be well below $0.1$\,events/day.
$^{12}$B events are produced from $^{12}$C in the detector either through an ($n$, $p$) reaction with spallation neutrons or a ($\mu^{-}, \nu_{\mu}$) reaction with cosmic muons, and then $\beta^{-}$ decay with a lifetime of 29.1\,ms and a $Q$-value of 13.4\,MeV.
Two $^{12}$B decays occurring one after the other or a combination of spallation neutron capture and a $^{12}$B decay could produce a background.
The rate of such background is evaluated using off-time windows to be $<0.03$\,events/day.

\begin{table}[ht]
\begin{center}
	\begin{tabular}{|c|c|c|}
	\hline
	Background & Rate (d$^{-1}$)& Gd-III/Gd-II\\
	\hline
	$^{9}$Li+$^{8}$He & $0.97^{+0.41}_{-0.16}$ & 0.78\\
	Fast-n + stop-$\mu$ & $0.604 \pm 0.051$ & 0.52 \\
	Accidental & $0.070 \pm 0.003$ & 0.27 \\
         $^{13}$C($\alpha$, n)$^{16}$O reaction & $< 0.1$ & not reported in Gd-II \\
         $^{12}$B & $< 0.03$ & not reported in Gd-II \\
	\hline
	\end{tabular}
	\caption{A summary of background rate estimations. Gd-III/Gd-II represents the reduction of the background rate with respect to the previous publication~\cite{ref:DCII_nGd} after scaling to account for the different prompt energy windows.}
	\label{table:BGSummary}
\end{center}
\end{table}

\section{Reactor-off Measurement}
\label{section:ReactorOff}
\begin{figure}
  \begin{center}
    \includegraphics[width=8cm]{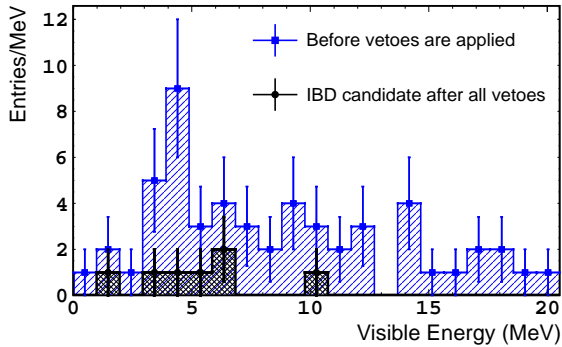}
    \caption{The prompt energy spectrum of IBD candidates observed in reactor-off running before background vetoes are applied (blue squares) and the spectrum of those after all vetoes are applied (black points).}
    \label{fig:ROFF}
 \end{center}
\end{figure}

Double Chooz collected 7.24 days of data with all reactors off in 2011 and 2012, in which background is dominant although a small contamination of residual reactor $\bar\nu_{e}$ is expected.
The number of residual reactor $\bar\nu_{e}$ is evaluated by a dedicated simulation study~\cite{ref:DC_Off} to be $1.57\pm0.47$ events.
54 events are selected by the delayed coincidence in the reactor-off running, and among these, 7 events remain after all background vetoes are applied.
Figure~\ref{fig:ROFF} shows the energy spectrum of the prompt signal before and after all background vetoes are applied.
The prediction for the reactor-off running is given as a sum of the background and residual $\bar\nu_{e}$'s to be $12.9 ^{+3.1}_{-1.4}$.
The compatibility of the observed number of events to the prediction is 9.0\,\% (1.7\,$\sigma$).
This data set is used not only to validate the background estimation but also to constrain the total background rate in the neutrino oscillation analyses. 

\section{Neutrino Oscillation Analysis}
\label{section:Oscillation}
The number of observed IBD candidates, the prediction of the unoscillated reactor neutrino signal and the estimated background contaminations are summarized in Table~\ref{table:IBDcandidates}.
In 460.67 days, 17351 IBD candidates are observed in reactor-on running, whereas the prediction including the background is $18290^{+370}_{-330}$ in absence of neutrino oscillation.
Uncertainties on the signal and background normalization are summarized in Table~\ref{table:RateError}.
The deficit of the IBD candidates can be interpreted as a consequence of reactor neutrino oscillation.
In order to evaluate the consistency of the observed data with the prediction of neutrino oscillation and extract the value of the neutrino mixing angle $\theta_{13}$, $\chi^{2}$ tests are carried out assuming two flavor oscillation expressed by Eq.~\ref{eq:oscillation}, in which $\Delta m_{31}^{2}$  is taken from the MINOS experiment as $\Delta m_{31}^{2} = 2.44^{+0.09}_{-0.10} \times 10^{-3} {\rm eV}^2$ assuming normal hierarchy~\cite{ref:MINOS_dm2} (a consistent value is reported by the T2K experiment~\cite{ref:T2K_dm2}).
Two complementary analysis methods, referred to as {\it Reactor Rate Modulation} ({\it RRM}) and {\it Rate+Shape} ({\it R+S}) analyses, are performed.
The {\it RRM} analysis is based on a fit to the observed IBD candidate rate as a function of the prediction, which depends on the number of operating reactor cores and their thermal power~\cite{ref:DCII_RRM}.
The {\it Rate+Shape} analysis is based on a fit to the observed energy spectrum in which both the rate of IBD candidates and the spectral shape information are utilized to give constraints on systematic uncertainties and $\theta_{13}$.

\begin{table}[ht]
\begin{center}
	\begin{tabular}{|c|c|c|}
	\hline
	& Reactor On& Reactor Off\\
	\hline
	Live-time (days) & 460.67 & 7.24\\
	\hline
	IBD Candidates & 17351  & 7\\
	\hline
	\hline
	Reactor  $\bar\nu_{e}$ & $17530 \pm 320$ & $1.57\pm0.47$\\
	Cosomogenic $^{9}$Li/$^{8}$He & $447^{+189}_{-74}$ & $7.0^{+3.0}_{-1.2}$\\
	Fast-$n$ and stop-$\mu$ & $278\pm23$ & $3.83\pm0.64$\\
	Accidental BG & $32.3 \pm 1.2$ & $0.508\pm0.019$ \\
	\hline
	Total Prediction & $18290^{+370}_{-330}$ & $12.9^{+3.1}_{-1.4}$ \\
	\hline
	\end{tabular}
	\caption{Summary of observed IBD candidates with the prediction of the unoscillated reactor neutrino signal and background. Neutrino oscillation is not included in the prediction.}
	\label{table:IBDcandidates}
\end{center}
\end{table}

\begin{table}[ht]
\begin{center}
	\begin{tabular}{|c|c|c|}
	\hline
	Source & Uncertainty (\%) & Gd-III/Gd-II\\
	\hline
	Reactor flux & 1.7 & 1.0\\
	Detection efficiency & 0.6 & 0.6 \\
	$^{9}$Li + $^{8}$He BG & $+1.1$ / $-0.4$ & 0.5\\
	Fast-n and stop-$\mu$ BG & 0.1 & 0.2\\
	Statistics & 0.8 & 0.7\\
	\hline
	Total & $+ 2.3$ / $-2.0$ & 0.8 \\
	\hline
	\end{tabular}
	\caption{Summary of signal and background normalization uncertainties relative to the signal prediction. The statistical uncertainty is calculated as a square root of the predicted number of IBD signal events. Gd-III/Gd-II represents the reduction of the uncertainty with respect to the previous publication~\cite{ref:DCII_nGd}.}
	\label{table:RateError}
\end{center}
\end{table}

\subsection{Reactor Rate Modulation Analysis}
\label{section:RRM}
\begin{figure}
  \begin{center}
    \includegraphics[width=9cm]{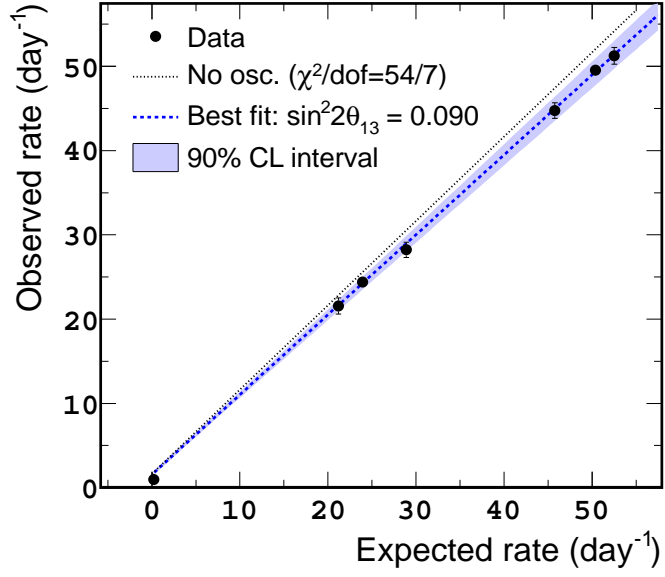}
    \caption{Points show the correlation between the expected and observed rates for different reactor powers. The first point refers to the reactor-off data. Overlaid lines are the prediction from the null oscillation hypothesis and the best {\it RRM} fit. In this fit, the background rate is constrained by the uncertainty on its estimation.}
    \label{fig:RRMfit}
 \end{center}
\end{figure}

The neutrino mixing angle $\theta_{13}$ can be determined from a comparison of the observed rate of IBD candidates ($R^{\rm obs}$) with the expected one ($R^{\rm exp}$) for different reactor power conditions.
In Double Chooz, there are three well defined reactor configurations: 1) two reactors are on (referred to as 2-On); 2) one of the reactors is off (1-Off); and 3) both reactors are off (2-Off).
By comparing $R^{\rm obs}$ at different reactor powers to the corresponding $R^{\rm exp}$, $\theta_{13}$ and the total background rate ($B$) are simultaneously extracted from the linear correlation between $R^{\rm obs}$ and $R^{\rm exp}$.
In the {\it RRM} analysis, the data set is divided into seven bins by the reactor thermal power ($P_{\rm th}$) conditions: one bin in the 2-Off period; three bins with mostly 1-Off; and three bins with 2-On (see Fig.~\ref{fig:RRMfit}). 

Three sources of systematic uncertainties on the IBD rate are considered in the {\it RRM} analysis: IBD signal detection efficiency ($\sigma_{\rm{d}}$=0.6\%); residual $\bar\nu_{e}$ prediction ($\sigma_{{\nu}}$=30\%); and prediction of the reactor flux in reactor-on data ($\sigma_{\rm{r}}$). 
$\sigma_{\rm{r}}$ depends on the reactor power and it ranges from 1.73\,\% at full reactor power to 1.91\,\% when one or two reactors are not at full power.

$\chi^{2}$ of the {\it RRM} fit is defined as follows:
\begin{eqnarray}
\label{eq:chi2RRM}
\chi^{2} &=& \chi^{2}_{\rm on} + \chi^{2}_{\rm off} + \chi^{2}_{\rm bg} +  \frac{\epsilon_{\rm d}^{2}}{\sigma_{\rm d}^{2}} + \frac{\epsilon_{\rm r}^{2}}{\sigma_{\rm r}^{2}} +  \frac{\epsilon_{\rm \nu}^{2}}{\sigma_{\rm \nu}^{2}}\\
\label{eq:chi2RRM_on}
\chi^{2}_{\rm on} &=& \displaystyle \sum_{i=1}^{\rm 6} \frac{\left(R_{i}^{\rm obs} - R_{i}^{\rm exp}-B\right)^2}{ (\sigma_{i}^{\rm stat})^2} \\
\label{eq:chi2RRM_off}
\displaystyle \chi^{2}_{\rm off} &=& 2 \left[ N_{\rm off}^{\rm obs} \ln \left( \frac{N_{\rm off}^{\rm obs}}{N_{\rm off}^{\rm exp}} \right) + N_{\rm off}^{\rm exp} - N_{\rm off}^{\rm obs}\right] \\
\label{eq:chi2RRM_bg}
\displaystyle \chi^{2}_{\rm bg} &=& \frac{\left(B-B^{\rm exp}\right)^2}{\sigma_{\rm bg}^2},
\end{eqnarray}
where the expected rate $R_{i}^{\rm exp}$ is varied to account for the systematic effects as a function of the parameters $\epsilon_{x}$ in the fit.
Neutrino oscillation is also accounted for in $R_{i}^{\rm exp}$.
$\sigma_{i}^{\rm stat}$ is the statistical uncertainty on the rate measurement.
The last three terms in Eq.~\ref{eq:chi2RRM} apply the constraints to the fit parameters from the estimated systematic uncertainties.
The systematic uncertainty on the reactor flux prediction is considered to be correlated between bins as the dominant source is the production cross-section measured by Bugey4~\cite{ref:Bugey4} which is independent of the thermal power. 
This is a conservative approach for the $\sin^{2}2\theta_{13}$ measurement.
$\chi^{2}_{\rm off}$ represents the contribution from the 2-Off data, in which $N_{\rm off}^{\rm obs}$ and $N_{\rm off}^{\rm exp}$ are the observed and expected number of IBD candidates.
$N_{\rm off}^{\rm exp}$ is given by the sum of the residual $\bar\nu_{e}$'s and the background.
A constraint to the total background rate is given by $\chi^{2}_{\rm bg}$.
The prediction of the total background rate and its uncertainty ($\sigma_{\rm bg}$) are given as: $B^{\rm exp} = 1.64^{+0.41}_{-0.17}$ events/day (see Section~\ref{section:Background}).

A $\chi^{2}$ scan of $\sin^{2}2\theta_{13}$ is carried out minimizing it with respect to the total background rate and three systematic uncertainty parameters for each value of $\sin^{2}2\theta_{13}$.
The best-fit gives $\sin^{2}2\theta_{13} = 0.090^{+0.034}_{-0.035}$ where the uncertainty is given as the range of $\chi^{2} < \chi^{2}_{\rm min} + 1.0$ with $\chi^{2}_{\rm min}/d.o.f.=4.2/6$.
The total background rate is found to be $B=1.56^{+0.18}_{-0.16}$\,events/day from the output of the fit.
Figure~\ref{fig:RRMfit} shows the correlation of the expected and observed IBD candidate rate with the best-fit prediction.

\begin{figure}
  \begin{center}
    \includegraphics[width=9cm]{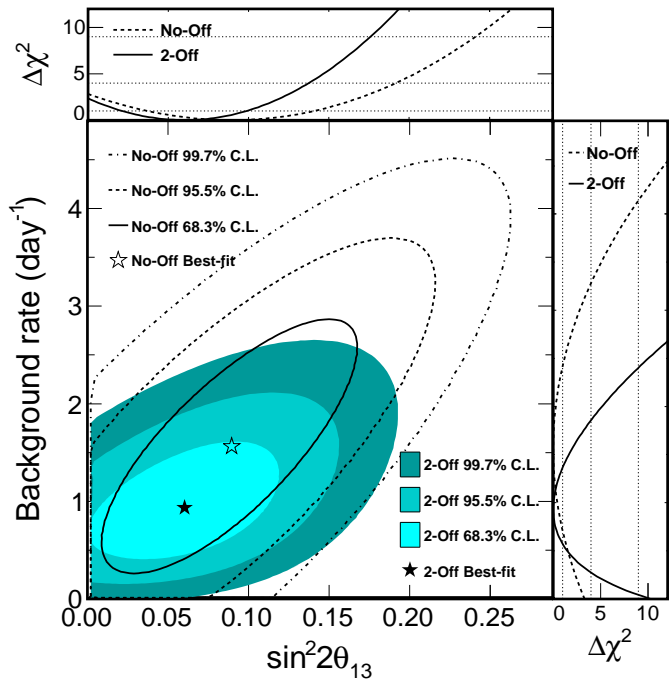}
    \caption{68.3, 95.5 and 99.7\,\% C.L. allowed regions on ($\sin^{2}2\theta_{13}$, $B$) plane obtained by the {\it RRM} fit with 2-Off data (colored contours). Overlaid contours (black lines) are obtained without the 2-Off data.  Background rate is not constrained by the estimation in both cases.}
    \label{fig:RRMContour}
  \end{center}
\end{figure}

The {\it RRM} fit is carried out with different configurations for validation.
First, the constraint on the total background rate ($\chi^{2}_{\rm bg}$) is removed, by which $B$ is treated as a free parameter in the fit.
This provides a cross-check and a background model independent measurement of $\theta_{13}$.
A global scan is carried out on the ($\sin^{2}2\theta_{13}$, $B$) grid minimizing $\chi^{2}$ at each point with respect to the three systematic uncertainty parameters.
The minimum $\chi^{2}$, $\chi^{2}_{\rm min}/d.o.f.=1.9/5$, is found at $\sin^{2}2\theta_{13} = 0.060\pm0.039$ and $B=0.93^{+0.43}_{-0.36}$\,events/day.
The value of $\sin^{2}2\theta_{13}$ is consistent with the {\it RRM} fit with background constraint.
Next, the reactor-off term ($\chi^{2}_{\rm off}$) is removed (constraint on the background is still removed in this case).
This configuration tests the impact of the data in reactor-off running to the precision of $\theta_{13}$ measurement.
The best fit without the 2-Off data is obtained with $\sin^{2}2\theta_{13} = 0.089\pm0.052$ and $B=1.56\pm0.86$\,events/day where $\chi^{2}_{\rm min}/d.o.f = 1.3/4$.
Figure~\ref{fig:RRMContour} shows the allowed parameter space on the ($\sin^{2}2\theta_{13}$, $B$) plane obtained by the {\it RRM} fit with and without the 2-Off data.
The precision of $\sin^{2}2\theta_{13}$ is significantly improved with the constraint on the total background rate given by the reactor-off measurement, which is a unique feature of Double Chooz with just two reactors.

\subsection{Rate + Shape Analysis}
\label{section:RS}
The {\it Rate+Shape} analysis is based on a comparison of the energy spectrum between the observed IBD candidates and the prediction.
The value of $\chi^{2}$ in the {\it R+S} fit is defined as follows:
\begin{eqnarray}
  \label{eq:chi2def}
  \chi^{2}   &=   \displaystyle \sum_{i=1}^{40} \sum_{j=1}^{40} (N_{i}^{\rm obs} - N_{i}^{\rm exp}) M_{ij}^{-1} (N_{j}^{\rm obs} - N_{j}^{\rm exp})
    + \sum_{k=1}^{5} \frac{\epsilon_{k}^{2}}{\sigma_{k}^{2}} \nonumber \\
    & \displaystyle+ (\epsilon_{a}, \epsilon_{b}, \epsilon_{c}) 
    \begin{pmatrix} \sigma_{a}^2 & \rho_{ab} \sigma_{a} \sigma_{b} & \rho_{ac} \sigma_{a} \sigma_{c} \\
    \rho_{ab} \sigma_{a} \sigma_{b} & \sigma_{b}^2 & \rho_{bc} \sigma_{b} \sigma_{c} \\
   \rho_{ac} \sigma_{a} \sigma_{c} & \rho_{bc} \sigma_{b} \sigma_{c} & \sigma_{c}^2
  \end{pmatrix} ^{-1}
  \begin{pmatrix} \epsilon_{a} \\ \epsilon_{b} \\ \epsilon_{c}  \end{pmatrix} \nonumber \\
   &\displaystyle + 2 \left[ N_{\rm off}^{\rm obs} \cdot \ln \left( \frac{N_{\rm off}^{\rm obs}}{N_{\rm off}^{\rm exp}} \right) + N_{\rm off}^{\rm exp} - N_{\rm off}^{\rm obs} \right].
\end{eqnarray}
In the first term, $N_{i}^{\rm obs}$ and $N_{i}^{\rm exp}$ refer to the observed and expected number of IBD candidates in the $i$-th energy bin, respectively.
Neutrino oscillation is accounted for in $N_{i}^{\rm exp}$ by Eq.~\ref{eq:oscillation}.
Data are divided into 40 energy bins suitably spaced between 0.5 and 20\,MeV to examine the oscillatory signature given as a function of  $E_{\nu}/L$ and statistically separate the reactor $\bar\nu_{e}$ signals from the background by the different spectral shapes.
$M_{ij}$ is a covariance matrix to account for statistical and systematic uncertainties in each bin and the bin-to-bin correlations.
$M_{ij}$ consists of the following matrices:
\begin{equation}
\label{eq:covariance}
M_{ij} = M_{ij}^{\rm stat} + M_{ij}^{\rm flux} + M_{ij}^{\rm eff} + M_{ij}^{\rm Li/He(shape)} + M_{ij}^{\rm acc(stat)},
\end{equation}
where $M_{ij}^{\rm stat}$ and $M_{ij}^{\rm acc(stat)}$  are diagonal matrices for the statistical uncertainty of the IBD candidates and statistical component of the uncertainty of the accidental background rate;
$M_{ij}^{\rm flux}$ accounts for the uncertainty on the reactor $\bar\nu_{e}$ flux prediction;
$M_{ij}^{\rm eff}$ is given as $M_{ij}^{\rm eff} = \sigma_{\rm eff}^{2} N_{i}^{\rm exp} N_{j}^{\rm exp}$ where $\sigma_{\rm eff} = 0.6\,\%$ represents the uncertainty on the MC normalization summarized in Table~\ref{table:MCCorrection};
and $M_{ij}^{\rm Li/He(shape)}$  encodes the shape error in the measured $^{9}$Li+$^{8}$He spectrum.

$N_{i}^{\rm exp}$ is corrected for systematic effects in the fit with eight parameters ($\epsilon_{x}$).
Variations of $\epsilon_{x}$ are constrained by the second and third terms in Eq.~\ref{eq:chi2def} with the estimated uncertainties ($\sigma_{x}$).
The following systematic uncertainties are considered in addition to those accounted for by the covariance matrices: 
the uncertainty on $\Delta m_{31}^{2}$ ($\Delta m_{31}^{2} = 2.44^{+0.09}_{-0.10} \times 10^{-3} {\rm eV}^2$); 
the uncertainty on the number of residual $\bar\nu_{e}$'s in reactor-off running ($1.57\pm0.47$ events); 
two uncertainties on the $^{9}$Li + $^{8}$He and fast neutron + stopping muon background rates; 
the systematic component of the uncertainty on the accidental background rate (see Section~\ref{section:Background}); 
and uncertainties on the energy scale represented by three parameters.
The fit parameter for the accidental background rate is constrained only by the systematic component of the uncertainty, as it is fully bin-to-bin correlated, while the statistical part is bin-to-bin uncorrelated and therefore accounted for by the covariance matrix ($M_{ij}^{\rm acc(stat)}$).
Correction for the systematic uncertainty on the energy scale is given by a second-order polynomial as: $\delta(E_{\rm vis}) = \epsilon_{a} + \epsilon_{b} \cdot E_{\rm vis} + \epsilon_{c} \cdot E_{\rm vis}^{2}$, where $\delta(E_{\rm vis})$ refers to the variation of visible energy.
Uncertainties on $\epsilon_{a}$, $\epsilon_{b}$ and $\epsilon_{c}$ are given as $\sigma_{a} = 0.006\,{\rm MeV}$, $\sigma_{b} = 0.008$ and $\sigma_{c} = 0.0006\,{\rm MeV}^{-1}$, respectively, taking into account those listed in Table~\ref{table:EnergyUncertainty}.
Constraints on $\epsilon_{a}$, $\epsilon_{b}$ and $\epsilon_{c}$ are given by a matrix in which correlations are taken into account with the following parameters: $\rho_{ab} = -0.30$, $\rho_{bc} = -0.29$ and $\rho_{ac} = 7.1 \times 10^{-3}$.

The last term in the $\chi^{2}$ definition represents the contribution to the $\chi^{2}$ from the reactor-off running.
As the statistics in the reactor-off running is low, only the number of IBD candidates ($N^{\rm obs}_{\rm off}$) is compared with the prediction ($N^{\rm exp}_{\rm off}$) by a log-likelihood based on Poisson statistics.

\begin{figure}
  \begin{center}
    \includegraphics[width=10cm]{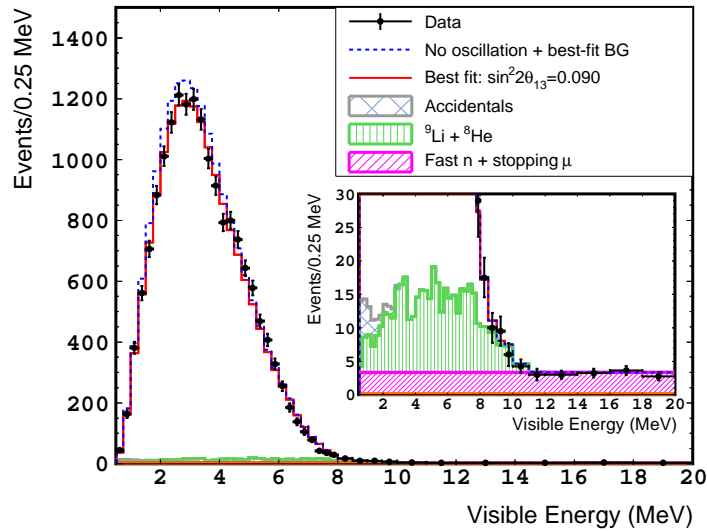}
    \caption{The measured energy spectrum of the prompt signal (black points) superimposed on the prediction without neutrino oscillation (blue dashed line) and the best-fit with $\sin^{2}2\theta_{13} = 0.090$ (red line). Background components after the fit are also shown with different colors: accidental (grey, cross-hatched); $^{9}$Li+$^{8}$He (green, vertical-hatched); and fast neutron + stopping muons (magenta, slant-hatched).}
    \label{fig:Spectrum}
  \end{center}
\end{figure}

A scan of $\chi^{2}$ is carried out over a wide range of $\sin^{2}2\theta_{13}$, minimizing it with respect to the eight fit parameters for each value of $\sin^{2}2\theta_{13}$.
The minimum $\chi^{2}$ value, $\chi^{2}_{\rm min}/d.o.f. = 52.2/40$, is found at $\sin^{2}2\theta_{13} = 0.090^{+0.032}_{-0.029}$, where the error is given as the range which gives $\chi^{2} < \chi^{2}_{\rm min} + 1.0$.
Best-fit values of the fit parameters are summarized in Table~\ref{table:BestFit} together with the input values and the uncertainties.
Figure~\ref{fig:Spectrum} shows the energy spectrum of the prompt signal superimposed on the best-fit prediction and the background components.
Assuming the inverted hierarchy with $|\Delta m_{31}^{2}| = 2.38^{+0.09}_{-0.10} \times 10^{-3} {\rm eV}^2$~\cite{ref:MINOS_dm2}, the best-fit is found at $\sin^{2}2\theta_{13} = 0.092^{+0.033}_{-0.029}$ with $\chi^{2}_{\rm min}/d.o.f. = 52.2/40$.

\begin{table}[ht]
  \footnotesize
  \begin{center}
    \begin{tabular}{| l | c | c |}
      \hline
      Fit Parameter & Input Value & Best-Fit Value \\
      \hline
      Li+He bkg. (d$^{-1}$) & $0.97^{+0.41}_{-0.16}$ & $0.74\pm0.13$ \\
      Fast-n + stop-$\mu$ bkg. (d$^{-1}$) & $0.604\pm0.051$ & $0.568^{+0.038}_{-0.037}$ \\
      Accidental bkg. (d$^{-1}$) & $0.0701\pm 0.0026$ & $0.0703\pm0.0026$ \\
      Residual $\bar\nu_{e}$ & $1.57\pm0.47$ & $1.48\pm0.47$ \\
      $\Delta m^2\ (10^{-3}$ eV$^2$) & $2.44^{+0.09}_{-0.10}$ & $2.44^{+0.09}_{-0.10}$ \\
      E-scale $\epsilon_{a}$ & $0\pm0.006$ & $0.001^{+0.006}_{-0.005}$  \\
      E-scale $\epsilon_{b}$ & $0\pm0.008$ & $-0.001^{+0.004}_{-0.006}$ \\
      E-scale $\epsilon_{c}$ & $0\pm0.0006$ & $-0.0005^{+0.0007}_{-0.0005}$ \\
      \hline
   \end{tabular}
  \end{center}
  \caption{Input values of fit parameters with the estimated uncertainties. Best-fit values and their errors are the output of the $\it Rate+Shape$ fit.}
  \label{table:BestFit}
\end{table}

A cross-check of the {\it R+S} fit is carried out removing the constraint to fit parameters for the $^{9}$Li+$^{8}$He and correlated background rates.
The minimum $\chi^{2}$, $\chi^{2}_{\rm min}/d.o.f. = 46.9/38$, is found at $\sin^{2}2\theta_{13} = 0.088^{+0.030}_{-0.031}$ with $^{9}$Li+$^{8}$He rate of $0.49^{+0.16}_{-0.14}$\,events/day and correlated background rate of $0.541^{+0.052}_{-0.048}$\,events/day.
The error for each parameter is defined as the range of $\chi^{2} < \chi^{2}_{\rm min} + 1.0$.
A consistent value of $\sin^{2}2\theta_{13}$ is thus obtained without the constraint to the background rates and the size of the errors are comparable after the fit.
This indicates that the uncertainties on the background rates are strongly suppressed in the {\it R+S} fit by the spectral shape information, and the output value of $\theta_{13}$ is robust with respect to the background estimation.

As a further cross-check, $\theta_{13}$ is found to be $\sin^{2}2\theta_{13} = 0.090^{+0.036}_{-0.037}$ by a comparison of the total observed rate to the prediction ({\it Rate-only} fit).
Observed rates in the reactor-on and reactor-off periods are separately used in the fit.

\begin{figure}
  \begin{center}
    \includegraphics[width=9cm]{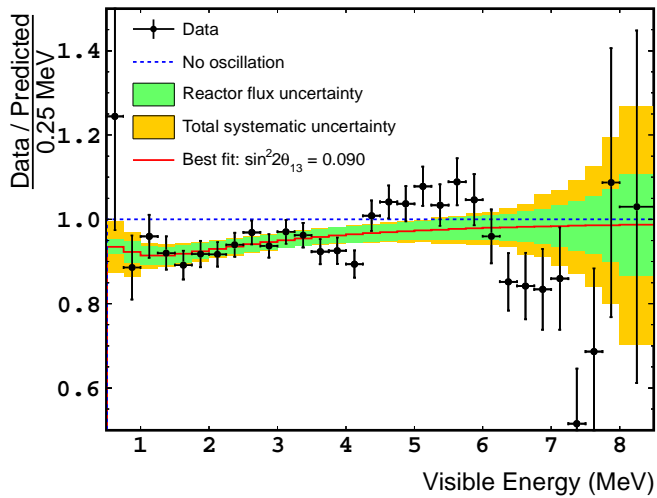}
    \caption{Black points show the ratio of the data, after subtraction of the background, to the non-oscillation prediction as a function of the visible energy of the prompt signal. Overlaid red line is the rate of the best-fit to the non-oscillation prediction with the reactor flux uncertainty (green) and total systematic uncertainty (orange).}
    \label{fig:SpectrumRatio}
  \end{center}
\end{figure}

Figure~\ref{fig:SpectrumRatio} shows the ratio of the data to the null oscillation prediction after subtraction of the background as a function of the visible energy of the prompt signal.
An energy dependent deficit is clearly seen in the data below 4\,MeV, which is consistent with the expectation from reactor neutrino oscillation.
On the other hand, besides the oscillatory signature, a spectrum distortion is observed at high energy above 4\,MeV, which can be characterized by an excess around 5\,MeV and a deficit around 7\,MeV.
In order to examine the impact of the excess on the measurement of $\theta_{13}$, a test of the {\it R+S} fit is carried out with an artificial excess in the prediction peaked at around 5\,MeV.
The normalization of the excess is left free in the test fit.
Among the outputs of the test fits with different peak energies and the widths of the excess, the maximum variations of $\sin^{2}2\theta_{13}$ and the output $^{9}$Li+$^{8}$He rate are within, respectively, 30\,\% and 10\,\% of their uncertainties.
With this, we conclude that the impact of the deviation in the observed energy spectrum on the $\sin^{2}2\theta_{13}$ measurement is not significant.
In addition, measured value of $\sin^{2}2\theta_{13}$ by the {\it R+S} fit agrees with that from {\it RRM} analysis independently of the spectrum shape, which demonstrates the robustness of the $\theta_{13}$ measurement despite the observed distortion.
Possible causes of the spectrum distortion are investigated in Section~\ref{section:Distortion}.

\subsection{Sensitivity with Near Detector}
\label{section:ProjectedSensitivity}
\begin{figure}
  \begin{center}
    \includegraphics[width=9cm]{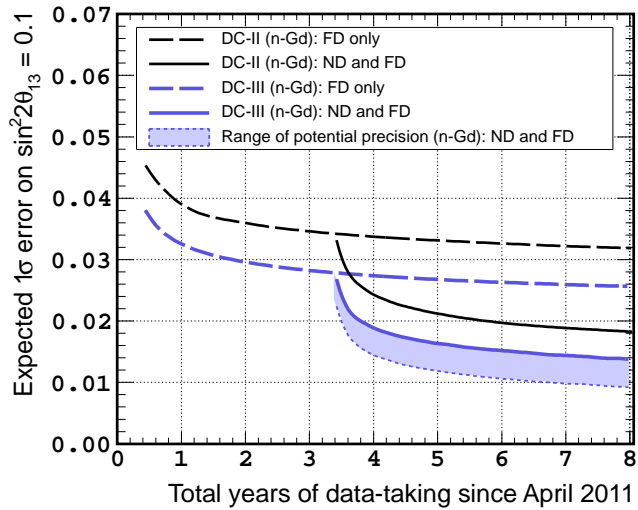}
    \caption{The projected sensitivity of Double Chooz with only the FD (blue dashed line) and that with the ND (blue solid line) based on the systematic uncertainties described in this paper. Assumptions on the relative uncertainties between the two detectors are described in text. Shaded region represents the range of improvements expected by the reduction in the systematic uncertainties and the lower edge corresponds to no systematic uncertainty besides the reactor flux. Overlaid black curves are the sensitivity based on the analysis reported in the previous publication~\cite{ref:DCII_nGd}. Only the IBD events with neutrons captured on Gd are used.}
    \label{fig:Sensitivity}
  \end{center}
\end{figure}

Figure~\ref{fig:Sensitivity} shows the projected sensitivity by the {\it R+S} fit with the ND based on the systematic uncertainties described in this paper.
We evaluated the following inputs for the sensitivity calculation: 
0.2\,\% uncertainty on the relative detection efficiency between the FD and ND ('IBD selection' in Table~\ref{table:MCCorrection} since all other contributions are expected to be suppressed);
the portion of the reactor flux uncertainty which is uncorrelated between the detectors is 0.1\,\% considering geometrical configuration of the Double Chooz sites;
background in the ND is estimated by scaling from the FD using measured muon fluxes at both detector sites.
The sensitivity curve is shown with the shaded region representing the range of improvements expected by the reduction in the systematic uncertainties (e.g. current systematic uncertainty on the background rate estimate is restricted by the statistics and therefore improvement on this is expected). 
The projected sensitivity with the ND reaches $\sigma(\sin^{2}2\theta_{13}) = 0.015$ in 3 years based on current knowledge and could be improved toward 0.010 with further analysis improvements.

An alternative curve in Fig.~\ref{fig:Sensitivity} shows the sensitivity based on the analysis reported in the previous publication~\cite{ref:DCII_nGd}.
One can conclude from the comparison that the improvement of the analysis described in this paper has a strong impact on the sensitivity of the future Double Chooz with the ND and the uncertainty on the $\sin^{2}2\theta_{13}$ is expected to be dominated by the statistical uncertainty even after 3 years with the improved analysis.

\section{Observed Spectrum Distortion}
\label{section:Distortion}
As is shown in Fig.~\ref{fig:SpectrumRatio}, a spectrum distortion is found above 4\,MeV of the prompt energy.
This trend is confirmed in the energy spectra reported in past publications: neutron captures on Gd and H by Double Chooz~\cite{ref:DCII_nGd, ref:DCII_nH} and neutron captures on Gd by the CHOOZ experiment~\cite{ref:CHOOZ}.
The observed IBD rates are higher than the prediction around 5\,MeV in these earlier publications although they are not significant enough to conclude the existence of an excess.
Without making any hypothesis on the overall shape of the distortion\footnote{During the final revision of this manuscript, a paper~\cite{ref:Dwyer} has been posted that tries to explain the possible origin of the excess in $4 < E_{\rm vis} < 6$\,MeV region.}, the bump around 5\,MeV has been selected as the major feature to estimate its significance.
The approaches used for this investigation are described in this section.

The energy scale around 5\,MeV is confirmed by spallation neutrons captured on carbon(C) which, due to the smaller capture cross section on C than on Gd, occur predominantly in the GC and result in an energy peak at 5\,MeV.
The energy scale of the C capture peaks agrees well, within 0.5\,\%, between the data and MC simulation.
Note that the energy resolution of the data is also in good agreement with that of the MC.
In addition, $\beta$ decays of $^{12}$B collected in the data are used to further test the energy scale as a cause of the excess.
No distortion is observed in the comparison of the energy spectrum of $^{12}$B between the data and MC simulation.

\begin{figure}
  \begin{center}
    \includegraphics[width=7.5cm]{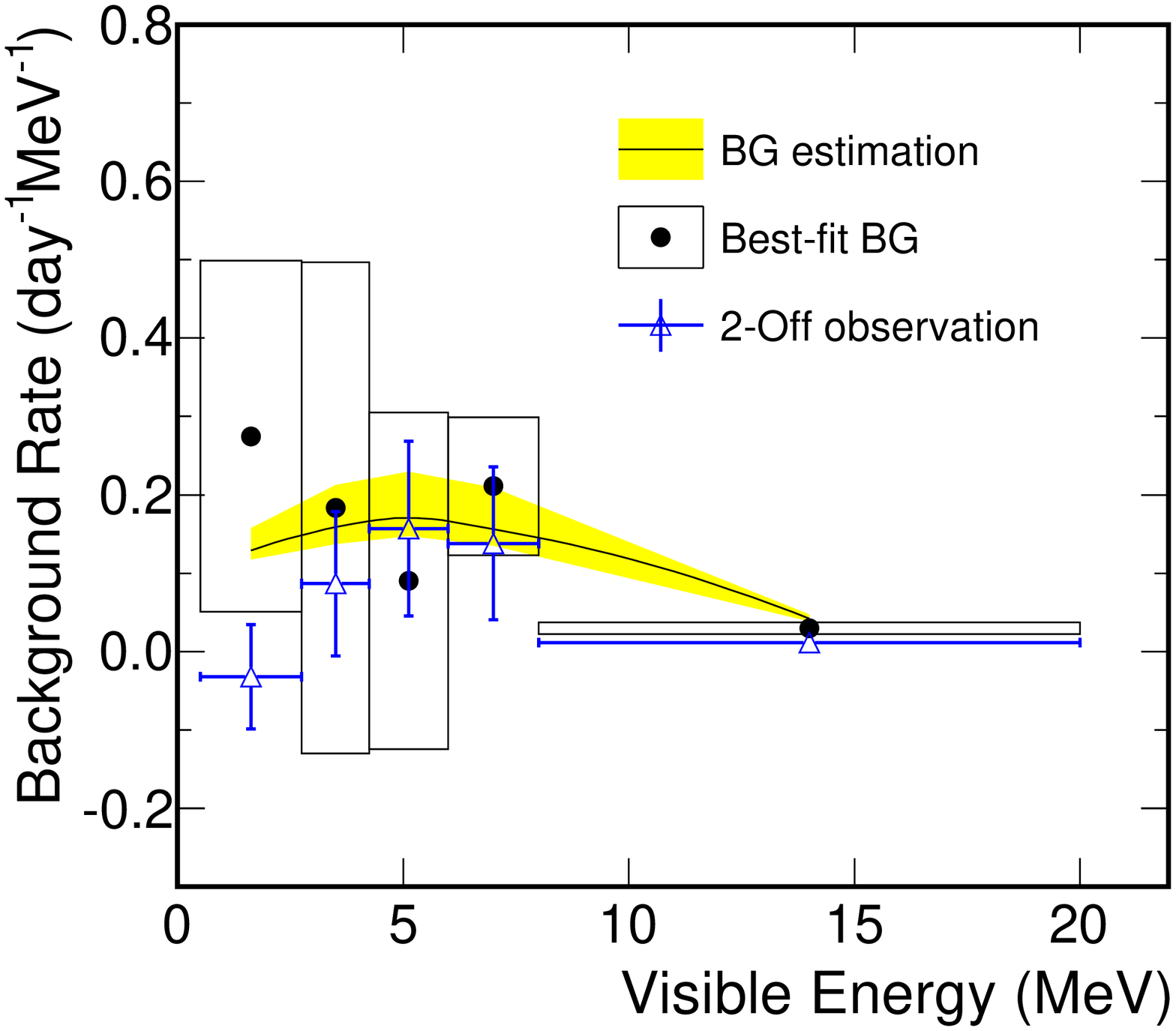}
    \includegraphics[width=7.5cm]{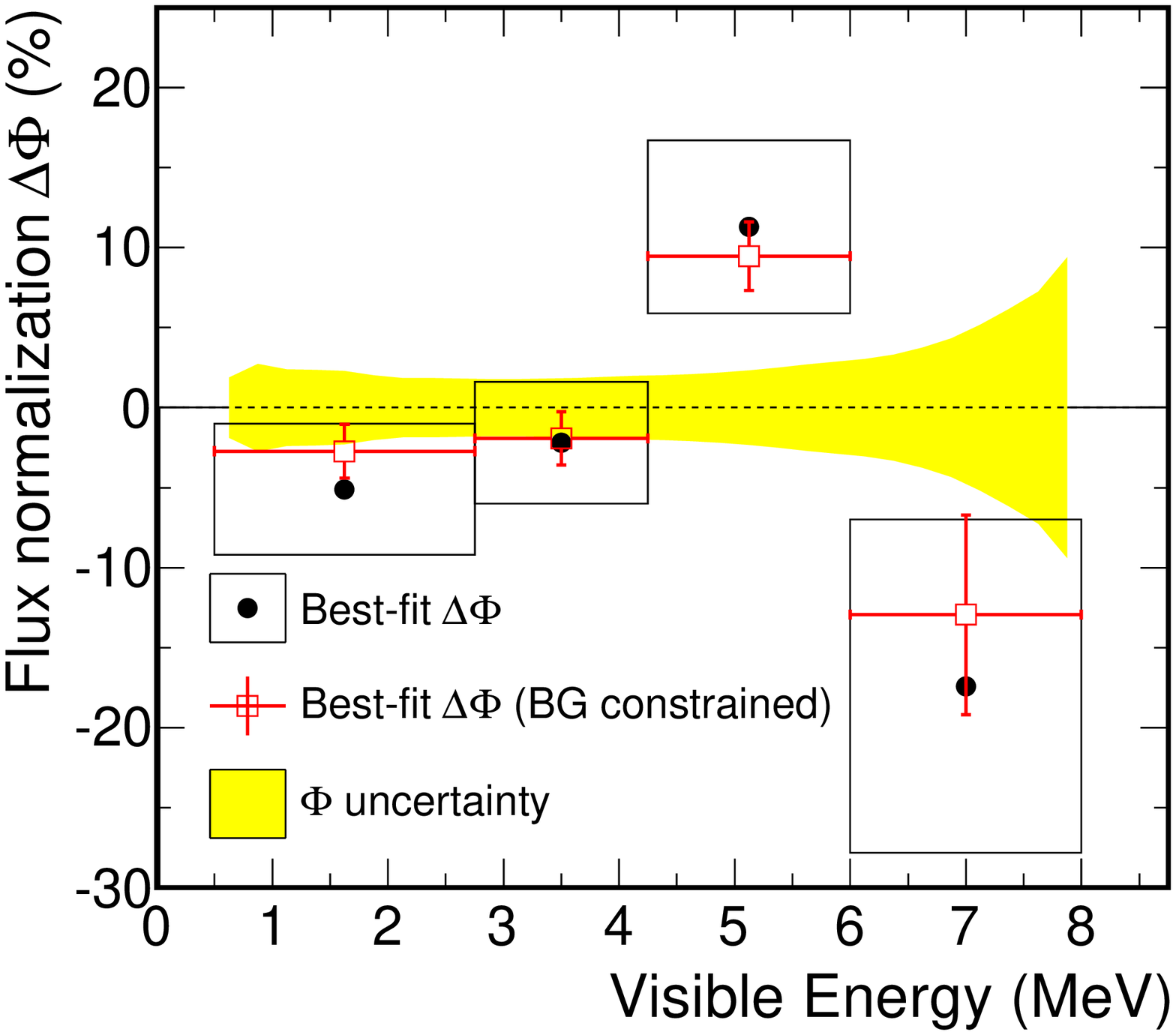}
    \caption{Output of the background rates and reactor flux normalizations from the independent {\it eRRM} fits for five energy regions with an additional constraint on $\sin^{2}2\theta_{13}$. The constraints to the reactor flux and background rate are removed in the fit. Left: Best-fit of background rates and the errors for the five data samples (black points and boxes) overlaid with the background rate estimation (line) and the observed rate in reactor-off running (blue empty triangles) with the uncertainties. Right: Black points and boxes show the best-fit of flux normalization with respect to the prediction and the error for the four data samples (background is dominant above 8\,MeV and therefore not sensitive to the reactor flux). Uncertainties on the background estimation and reactor flux prediction are shown by the yellow bands. Red empty squares show the best-fit and the error with the BG constraint from the estimations in the {\it eRRM} fit.}
    \label{fig:RRMtest}
  \end{center}
\end{figure}

\paragraph{Deviation from reactor flux prediction}
If the excess around 5\,MeV is due to unknown backgrounds, the rate of the excess should be independent of the reactor power, while if it is due to the reactor flux,  the rate of the excess should be proportional to the reactor power.
In order to evaluate the consistency of data with the reactor flux and background predictions, an energy-binned {\it RRM} fit ({\it eRRM} fit) is carried out with different configurations from Section~\ref{section:RRM}.
The data are divided into five samples by the visible energy of the prompt signal to investigate the energy dependence, where the energy binning is optimized to pick up excess around 5\,MeV.
The {\it eRRM} fit utilizes a correlation between the observed rate and reactor power and therefore is sensitive in distinguishing the background and reactor flux hypotheses as the cause of the excess.
First, constraints to the background rate and reactor flux are removed while a constraint to $\theta_{13}$ is given as $\sin^{2}2\theta_{13} = 0.090^{+0.009}_{-0.008}$ from the Daya Bay experiment~\cite{ref:DayaBay}.
Figure~\ref{fig:RRMtest} shows the best-fit of the background rate and relative normalization of the reactor flux for each energy range.
Background rates are consistent with the estimation and also consistent with the observed background rate in reactor-off running after subtracting the residual $\bar\nu_{e}$.
On the other hand, the output of the reactor flux normalization from the {\it eRRM} fit is higher by 2.0\,$\sigma$ than the prediction between 4.25 and 6\,MeV and lower by 1.5\,$\sigma$ between 6 and 8\,MeV.
The implication is that the observed spectrum distortion originates from the reactor flux prediction, while the unknown background hypothesis is not favored.

In order to evaluate the deviation from reactor flux prediction, we incorporate the background rate estimation as a constraint in the {\it eRRM} fit.
The significance of the excess and deficit in the flux prediction with respect to the systematic uncertainty reaches 3.0\,$\sigma$ and 1.6\,$\sigma$, respectively (red empty squares in Fig~\ref{fig:RRMtest}).  

\begin{figure}
  \begin{center}
    \includegraphics[width=9cm]{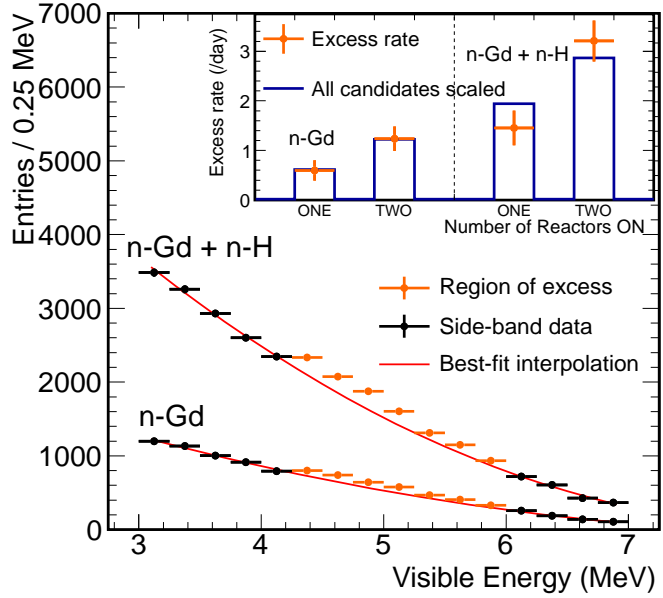}
    \caption{The energy spectrum of the prompt signal for IBD candidates with neutrons captured on Gd and one including H captures (Gd+H). Points show the data and lines show the second order polynomial functions. Inset figure: points show the correlations between the observed rate of the excess (defined in the text) and the number of operating reactors, and the histograms show the total IBD candidate rate (area normalized). The H capture sample includes accidental background with a rate comparable to the IBD signals and therefore the total rate of the Gd+H sample has an offset due to this background in addition to IBD signals which is proportional to the reactor power.}
    \label{fig:Correlation}
  \end{center}
\end{figure}

\paragraph{Correlation to reactor power}
Given the indication from the {\it eRRM} fit, the correlation between the rate of the excess and reactor power is further investigated by a dedicated study targeted on the region of the excess.
First, assuming the IBD rate is smoothly decreasing with the energy, its rate between 4.25 and 6\,MeV, where it is most enhanced, is estimated by an interpolation with a second order polynomial from the observed rate below 4.25\,MeV and above 6\,MeV as shown in Fig.~\ref{fig:Correlation}.
Second, the rate of excess is defined as the observed IBD candidate rate between 4.25 and 6\,MeV after subtracting the interpolation estimation, and the correlation between the rate of excess and the number of operating reactors is investigated.
If the excess is due to an unknown background, the rate of the excess should be independent of the reactor power, while as shown in the inset plot (left) in Fig.~\ref{fig:Correlation}, a strong correlation between the rate of excess and the number of operating reactors is confirmed.
The significance of the correlation becomes stronger by adding the IBD candidates with neutrons captured on H based on the same data set used in this paper and following the selection criteria described in Ref. \cite{ref:DCII_nH} (right-hand plot in the inset).

\section{Conclusion}
\label{section:Conclusion}
Improved measurements of the neutrino mixing angle $\theta_{13}$ have been performed by Double Chooz using two analysis methods,
based on the data corresponding to 467.90 days of live time.
A best-fit to the observed energy spectrum gives $\sin^{2}2\theta_{13} = 0.090^{+0.032}_{-0.029}$.
A consistent value of $\theta_{13}$,  $\sin^{2}2\theta_{13} = 0.090^{+0.034}_{-0.035}$, is obtained by a fit to the observed IBD rates in different reactor power conditions.
These two analyses utilize different information, energy spectrum shape and reactor rate modulation, to extract $\theta_{13}$, and therefore work as a cross-check to each other.

A spectrum distortion is observed at a high energy above 4\,MeV but its impact on the $\theta_{13}$ measurement is evaluated to be insignificant with respect to the uncertainty.
A strong correlation between the excess rate and the reactor power  is observed.
The significance of the excess between 4.25 and 6\,MeV including the uncertainty of the flux prediction is evaluated to be 3.0\,$\sigma$ assuming only standard IBD interactions.
In addition to the excess, a deficit is found between 6 and 8\,MeV with a significance of 1.6\,$\sigma$.

The near detector construction is nearing completion.  
As a consequence of the analysis improvements described in this paper, the projected sensitivity of Double Chooz reaches $\sigma(\sin^{2}2\theta_{13}) = 0.015$ in 3 years data taking with the ND, and could be further improved towards 0.010.

\acknowledgments

We thank the French electricity company EDF; the European fund FEDER;
the R\'egion de Champagne Ardenne; the D\'epartement des Ardennes;
and the Communaut\'e des Communes Ardennes Rives de Meuse.
We acknowledge the support of the CEA, CNRS/IN2P3, the computer center CCIN2P3, and LabEx UnivEarthS in France (ANR-11-IDEX-0005-02);
the Ministry of Education, Culture, Sports, Science and Technology of Japan (MEXT) and the Japan Society for the Promotion of Science (JSPS);
the Department of Energy and the National Science Foundation of the United States;
the Ministerio de Ciencia e Innovaci\'on (MICINN) of Spain;
the Max Planck Gesellschaft, and the Deutsche Forschungsgemeinschaft DFG (SBH WI 2152), the Transregional Collaborative Research Center TR27, the excellence cluster ``Origin and Structure of the Universe'', and the Maier-Leibnitz-Laboratorium Garching in Germany;
the Russian Academy of Science, the Kurchatov Institute and RFBR (the Russian Foundation for Basic Research);
the Brazilian Ministry of Science, Technology and Innovation (MCTI), the Financiadora de Estudos e Projetos (FINEP), the Conselho Nacional de Desenvolvimento Cient\'ifico e Tecnol\'ogico (CNPq), the S\~ao Paulo Research Foundation (FAPESP), and the Brazilian Network for High Energy Physics (RENAFAE) in Brazil.



\end{document}